# Breaking the mutual exclusivity between metallicity and ferroelectricity in a non-polar covalent semiconductor via orbital selective doping

Hui Li,[1] Yunfan Yang,[1,8] Junquan Huang,[2] Yukun Feng,[1] Guobin Wang,[3] Qinci Wu,[4] Jun Deng,[5] Zhaolong Liu,[1,8] Subi Du,[1,8] Dongliang Gong,[6] Zaihui Shen,[1,8] Anmin Nie,[2*] Yang Xu,[1,8] Junwei Yang,[7] Zesheng Zhang,[3] Huaping Song,[7] Jiangang Guo,[1,8] Wenjun Wang,[1,8] Hailin Peng,[4] Yongjun Tian,[2] and Xiaolong Chen*[1,8]

[1]Beijing National Laboratory for Condensed Matter Physics, Institute of Physics, Chinese Academy of Sciences, Beijing 100190, China

[2]Center for High Pressure Science, State Key Laboratory of Metastable Materials Science and Technology, Yanshan University, Qinhuangdao 066004, China

[3]Beijing Lattice Semiconductor Co., Ltd., Beijing 101300, China

[4]Center for Nanochemistry, Beijing Science and Engineering Center for Nanocarbons, Beijing National Laboratory for Molecular Sciences, College of Chemistry and Molecular Engineering, Peking University, Beijing, P. R. China.

[5]Center for High Pressure Science and Technology Advanced Research, Beijing, 100193, China.

[6]Institute of Electrical Engineering, Chinese Academy of Sciences, Beijing 100190, China

[7]PowerEpi Semiconductor Co., Ltd. Dongguan, Guangdong 523808, China

[8]School of Physical Sciences, University of Chinese Academy of Sciences, Beijing 100049, China

Corresponding authors: Anmin Nie, E-mail: anmin@ysu.edu.cn; Xiaolong Chen, E-mail: chenx29@iphy.ac.cn

**ABSTRACT:** The mutual exclusion of ferroelectricity and metallic conductivity is a long-standing tenet because itinerant electrons screen long-range Coulomb forces that stabilize the bulk polar order[1]. In 1965, Anderson and Blount proposed the existence of polar metals *via* structural phase transformation in metals[2], a prediction that was experimentally realized decades later in $LiOsO_3$[3]. However, it remains a purely structural analog that lacks macroscopically switchable ferroelectricity[4], prompting intensive efforts to focus on 2D layered metals like $WTe_2$[5], engineered heterostructures[6,7], and doped ferroelectric oxides[8,9,10]. The coexistence of switchable ferroelectricity and metallicity in bulk single crystals still remains a major challenge because of achieving robust polar distortion within a dense electron sea and penetrating a bulk metal with an external electric field to switch its polarization. Here, we break this paradigm by heavily doping a non-polar covalent semiconductor of cubic silicon carbide (3C-SiC) with nitrogen. This introduces heavy electron doping, inducing metallicity and driving a structural transition from the non-polar $F\bar{4}3$m to the polar $R3m$ symmetry via the pseudo-Jahn-Teller effect. Remarkably, we provide direct, atomic-scale visualization of about 180° polarization reversal under an external voltage bias in a ferroelectric metal. The strongly directional character of antibonding orbitals occupied by conduction electrons prevents them from screening the local Si-C polarization, resulting in the coexistence of metallicity and ferroelectricity. Ferroelectric tunnel junctions demonstrate nonvolatile memory properties with a well-defined high-resistance state (HRS) and low-resistance state (LRS), an ultrahigh response speed (~50 ns), an ultralow operating voltage (1 V), an endurance exceeding 85927 cycles, and a projected retention time of 100 years. Our results provide a novel strategy for pioneering ferroelectricity in a metal, a new ferroelectric metal platform for exploring exotic properties, and a ferroelectric device with high performance that meets the requirements for low consumption and high-speed non-volatile devices.

Ferroelectric materials, in which spontaneous polarization can be reversed under an external electric field, are fundamental to modern non-volatile memories, sensors and optoelectronic devices. Ferroelectricity (FE) is traditionally considered to emerge in insulators, where the absence of free carriers facilitates stable long-range dipolar ordering and hence robust FE. Conventional ferroelectric materials, however, have inherent shortcomings in speed, energy consumption, and interface bottlenecks due to their insulating nature.

Ferroelectric metals, which possess both metallicity and switchable polarization, offer exceptional promise for next-generation electronics thanks to simplified device geometry, ultra-high speed, and ultra-low power consumption. Frustratingly, free carriers in metals have long been thought to screen the long-range Coulomb interactions that stabilize spontaneous polarization, rendering metallicity and FE mutually exclusive in the same material[1]. In 1965, Anderson and Blount theorized that a metal could undergo a polar structural transition if the distortion were sufficiently decoupled from free electrons[2], spurring intense research efforts to search for or pioneer materials that simultaneously exhibit FE and metallicity[3-14]. Decades later, the observation of polar phase transitions in metallic $LiOsO_3$ provides compelling experimental validation of this theoretical prediction[3]. However, $LiOsO_3$ is not a real ferroelectric metal because its polarization cannot be switched by an external electric field[4].

Recently, researchers have turned to superlattices[6,7], doped-ferroelectric materials[8-10], and two-dimensional metallic systems[5,14], where FE and metallicity coexist through spatial or electronic decoupling, or through strong electron-phonon coupling (EPC). Nevertheless, these platforms either rely on pre-existing FE, interfacial or dimensional effects, or geometries in which field penetration differs fundamentally from that in a bulk metal.

Bulk single crystals provide the most stringent test of intrinsic ferroelectric metallicity because they avoid substrate clamping, interfacial dead layers, and dimensional confinement[1,15]. They also pose the hardest electrostatic problem: free carriers in three dimensions screen applied fields before the fields can penetrate deeply enough to reverse the polar lattice. Metallic conduction further complicates conventional macroscopic polarization measurements, making direct microscopic evidence of reversible polar displacement essential. Thus, a bulk single crystal that combines metallic conduction with electrically switchable polar order remains a critical missing case. However, bulk single crystals are pivotal for both fundamental understanding of ferroelectric metal and the realization of practical ferroelectric metal-based devices. Cubic silicon carbide (3C-SiC) provides an opportunity to address this problem. As a non-polar covalent wide-bandgap semiconductor that can be grown as wafer-scale single crystals and epitaxially grown on silicon substrates[16-24], 3C-SiC (space group: $F\bar{4}3m$) allows heavy nitrogen doping to introduce a high density of free electrons into the tetrahedral Si-C network and modify antibonding states[23,24,25], suggesting a possible route to destabilize the non-polar cubic lattice toward a polar distortion, alongside the metallicity. However, whether such doping can actually induce lattice distortion and thereby switchable FE in a metallic system has remained a long-standing open question.

Here, we resolve this long-standing question by achieving electrically switchable FE in heavily nitrogen (N)-doped 3C-SiC bulk single crystals with a metallic state. The high carrier density (~2.9 × $10^{20}$ $cm^{-3}$) yields a low room temperature resistivity of 0.52 mΩ·cm and drives a structural distortion from the parent non-polar $F\bar{4}3m$ phase to the polar $R3m$ (No. 160) phase. Using in-situ spherical aberration-corrected scanning transmission electron microscopy (STEM), we directly visualize about 180° polarization reversal under opposite biases, providing direct atomic-scale evidence of ferroelectric domain switching within a metallic lattice. Free carriers occupy antibonding orbitals, which weakens the screening of local Si–C polar distortions and thereby preserves ferroelectricity. Ferroelectric tunnel junctions directly fabricated on these ferroelectric metal single crystals demonstrate nonvolatile memory properties with a distinct high-resistive state (HRS) and a low-resistive state (LRS), an ultrahigh response speed (~50 ns), an ultralow operating voltage (~1 V), and excellent anti-fatigue and properties (more than 80000 cycles).

**Electrical and structural characterizations of N-doped 3C-SiC**

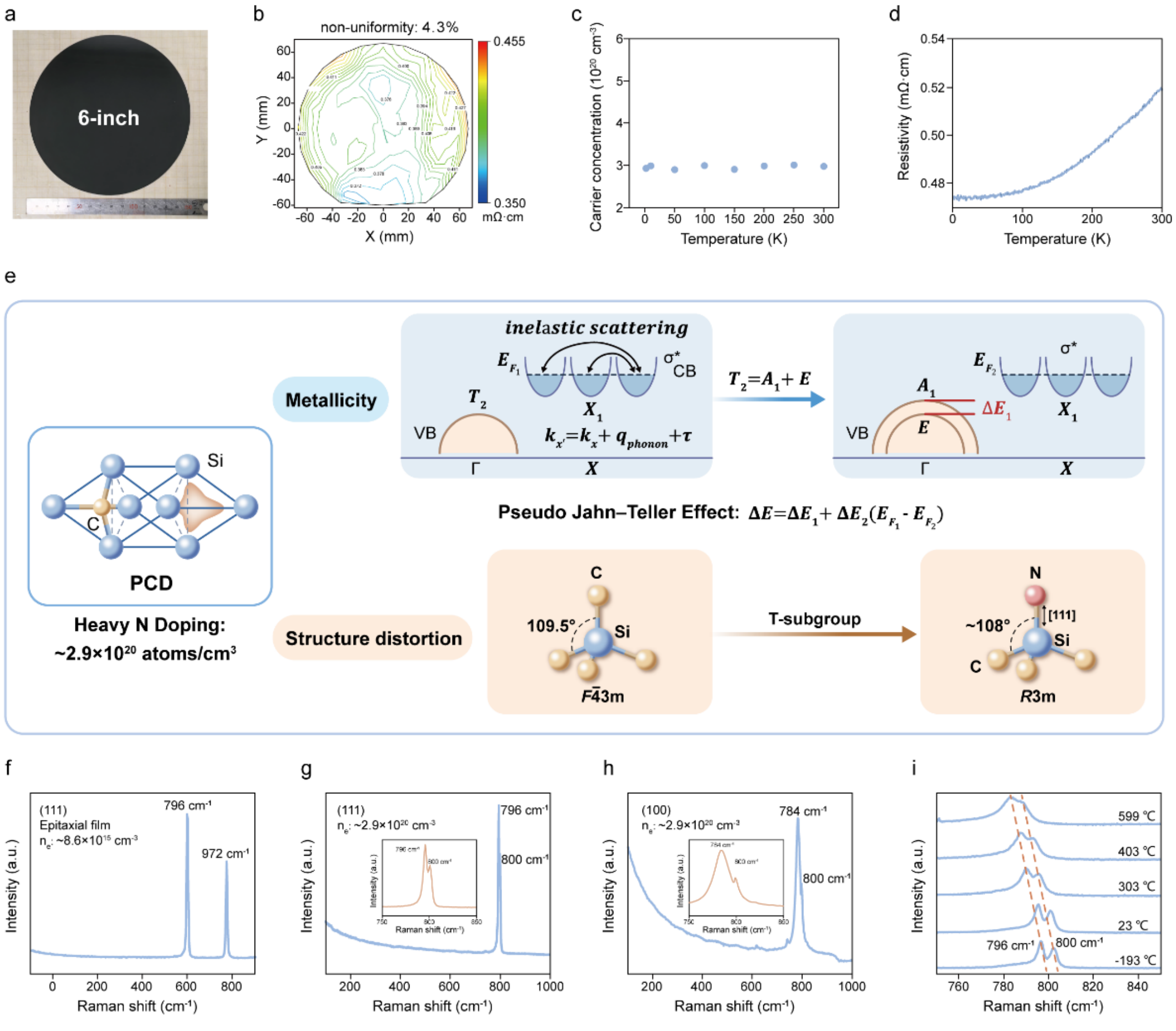

**Fig. 1 | Metallicity and lattice distortion of heavily N-doped 3C-SiC single crystals. a,** Photograph of a representative 6-inch single crystal wafer. **b,** Two-dimensional resistivity mapping of the 6-inch single crystal wafer, exhibiting good spatial uniformity. **c,** Carrier concentration from Hall measurements over 2-300 K. **d,** Temperature-dependent resistivity (ρ) measured from 2 to 300 K. The curve clearly demonstrates metallicity and strong EPC due to the slight temperature dependence of ρ. **e,** Illustration of the lattice distortion driven by the

pseudo Jahn-Teller effect (PJTE). $\Delta E_1$ and $\Delta E_2$ represent the changed $\sigma^*$ energy and electronic stabilization energy, respectively. CB and VB represent conduction band and valence band, respectively. The left inset shows the calculated partial charge density (PCD) of the conduction band minimum (CBM) with an iso-value of 0.035 e·Bohr$^{-3}$, revealing that electrons are isolated from the Si-C(N) bonds and occupy the $\sigma^*$ orbital. **f,** Raman spectrum of epitaxial 3C-SiC film (carrier concentration of ~8.6×10$^{15}$ cm$^{-3}$). The LO peak at 972 cm$^{-1}$ and the threefold degenerate TO peak at 796 cm$^{-1}$ are clearly seen, showing the cubic structure under low N doping. **g,h,** Raman spectra of heavily N-doped 3C-SiC single crystals measured on the (111) and (100) planes, respectively, showing the disappearance of the LO peak and the splitting of the threefold-degenerate TO peak (796 and 800 cm$^{-1}$ on (111) plane, 784 and 800 cm$^{-1}$ on (100)), evidencing strong EPC and $R3m$ lattice distortion. **i,** Temperature-dependent TO mode measured on the (111) plane from -193 °C to 599 °C. The splitting of the TO mode persists across the entire measured temperature range, indicating that the $R3m$ phase remains stable up to at least 599 °C. a.u., arbitrary units.

Atom probe tomography (APT) (Extended Data Fig. 1a), secondary ion mass spectrometry (SIMS) (Extended Data Fig. 1b), and electron energy loss spectroscopy (EELS) (Extended Data Fig. 1c-e) demonstrate a high N concentration (~2.9 × 10$^{20}$ atoms·cm$^{-3}$) with a uniform distribution in the 3C-SiC bulk single crystals. The homogeneous N distribution across the 6-inch wafer is corroborated by resistivity mapping (Fig. 1a-b). Hall measurements reveal a carrier concentration of ~2.9 × 10$^{20}$ cm$^{-3}$ (Fig. 1c), indicating nearly 100% ionization of N, in good agreement with our reported results[23,24]. The high-density carrier injection induces metallicity with a room-temperature resistivity of ~0.52 mΩ·cm (Fig. 1d) by shifting the Fermi level into the conduction band (CB) (Extended Data Fig. 2a,b)[23,25]. The calculated spatially resolved projected density of states (PDOS, Extended Data Fig. 2c) demonstrates that the non-zero density of states at the Fermi level predominantly originates from N, which serves as the primary origin of metallicity. The negative Hall coefficient (Extended Data Fig. 2d) reveals that electrons are the majority carriers. The carrier concentration and carrier mobility change little upon heating (Extended Data Fig. 2e).

The conduction electrons occupy the degenerate $\sigma^*$ antibonding orbital at the $X$ valleys of the Brillouin zone, as determined from the calculated partial charge density (PCD) of the CB minimum (CBM) (the inset of Fig. 1e), facilitating robust inter-valley electronic transitions. The corresponding momentum transfer wavevectors result in strong electron-phonon coupling (EPC), as revealed by the softened transverse optical (TO) phonons measured on the (100) planes (Fig. 1h). The disappearance of the longitudinal optical (LO) peak in the heavily N-doped 3C-SiC single crystals (Fig. 1f-h) shows strong EPC[31]. The substitution of C by N shortens the Si-C bond along the <111> axis and leads to carrier injection and thus the splitting of the non-degenerate $A_1$ singlet (Extended Data Fig. 2a,b). Thus, the pseudo-Jahn-Teller effect (PJTE) is triggered and induces the phase transformation from $F\bar{4}3m$ to $R3m$ (Fig. 1e)[26], which is the most energetically favorable distortion pathway (Supplementary text).

Undoped and lightly N-doped 3C-SiC ($n_e$:~8.6 × $10^{15}$ cm$^{-3}$, Supplementary Table 1) retains cubic symmetry, displaying a single TO Raman peak at 796 cm$^{-1}$ (Fig. 1f)[27]. In heavily N-doped samples, symmetry breaking splits the TO phonon into 800 cm$^{-1}$ and 796 cm$^{-1}$ measured on the (111) plane, 800 cm$^{-1}$ and 784 cm$^{-1}$ measured on the (100) plane (Fig. 1g-h)[23,24,27]. Raman mapping (Extended Data Fig. 3c-f) of the 796 cm$^{-1}$ peak measured on the (111) plane identified by electron backscatter diffraction (EBSD) (Extended Data Fig. 3a,b) demonstrates the single phase of the 6-inch wafer. Notably, the $R3m$ lattice distortion is quite stable, persisting up to 599 °C (the highest testing temperature of the equipment) as evidenced by the splitting of the TO peak (Fig. 1i).

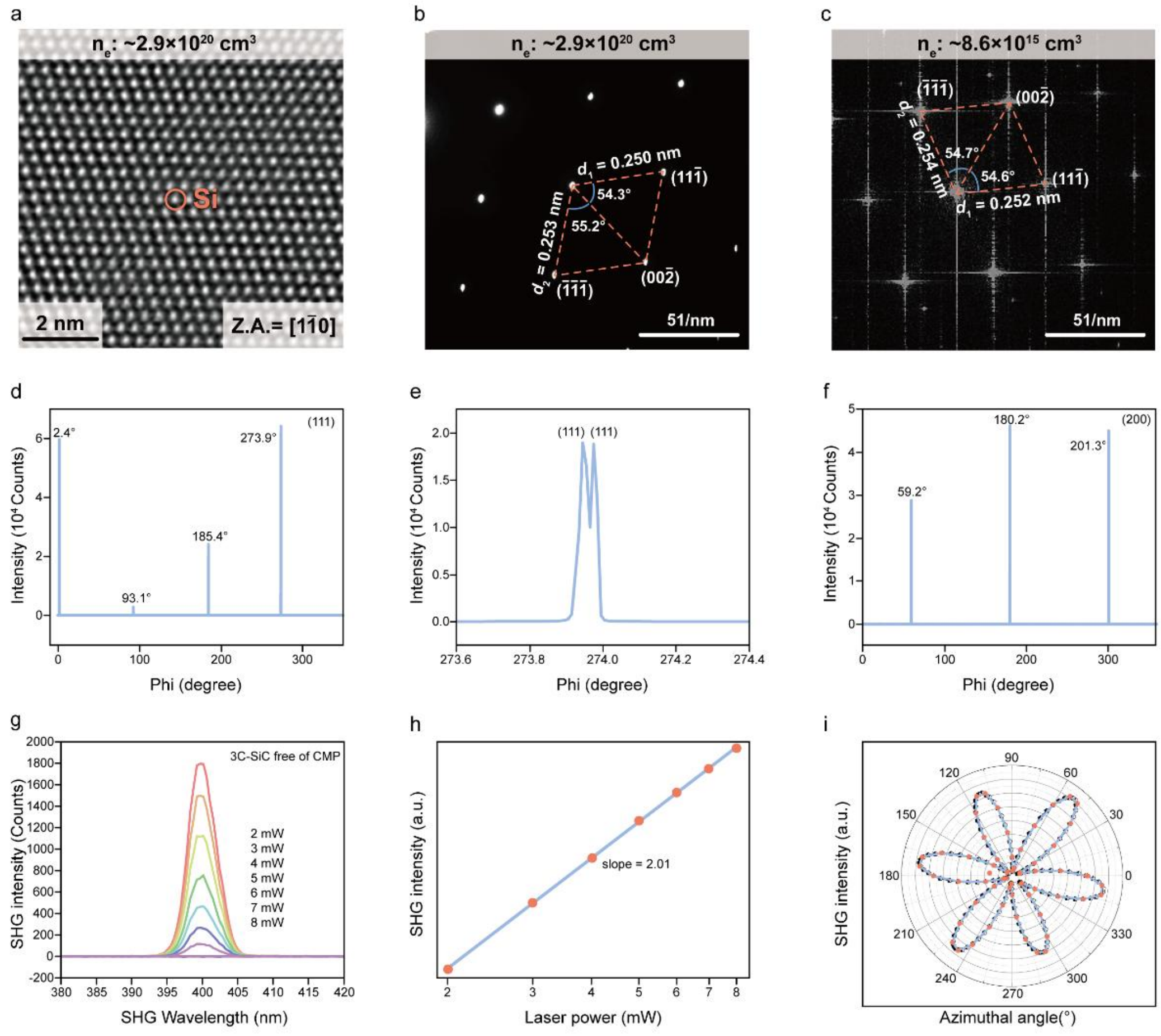

**Fig. 2 | HAADF-STEM, XRD, and SHG characterizations of heavily N-doped 3C-SiC single crystals. a,** An atomic-resolution HAADF-STEM image and (**b**) a SAED pattern along the $[1\bar{1}0]$ zone axis. The measured interplanar spacings $d_{(\bar{1}\bar{1}\bar{1})}$, $d_{(11\bar{1})}$, along with the interplanar angles between $(\bar{1}\bar{1}\bar{1})/(00\bar{2})$ and $(11\bar{1})/(00\bar{2})$ planes reveal the $R3m$ lattice distortion along the parent-cubic <111> direction upon heavily N doping. **c,** A FFT pattern for an epitaxial

3C-SiC film with a carrier concentration of ~$8.6\times10^{15}$ cm$^{-3}$. The nearly identical $d_{(\bar{1}1\bar{1})}$, $d_{(11\bar{1})}$, and the angles between the $(\bar{1}1\bar{1})/(00\bar{2})$ and $(11\bar{1})/(00\bar{2})$ planes show the cubic structure free of lattice distortion. **d,f,** XRD phi-scan profiles for (**d**) the (111) and (**f**) the (200) planes with out-of-plane orientations along the [100] and [111] directions, respectively, revealing the deviation of four-fold and three-fold azimuthal symmetries of {111} and {200} reflections. **e,** Enlargement of the (111) reflection portion around 273.9° shown in (**d**). The splitting of (111) reflections further substantiate lattice distortion. **g,** Power dependence of the SHG intensity excited by an 800 nm laser. **h,** A log-log plot of the SHG intensity versus laser power, revealing a linear relationship with a fitted slope of 2.01 and confirming the intrinsic SHG response. **i,** Polarization-angle-dependent SHG intensity excited by the 780 nm femtosecond laser, exhibiting a six-fold symmetric petal-like pattern. The result is in excellent agreement with the 3*m* point group, corroborating the *R*3*m* space group. Z.A., zone axis. a.u., arbitrary units.

The N-doped 3C-SiC single crystals are further characterized by the atomic-scale aberration-corrected high-resolution high angle annular dark field scanning transmission electron microscopy (HAADF-STEM) (Fig. 2a). In HAADF-STEM images, the atomic number (Z) contrast mechanism renders Si atoms with significantly higher intensity than C(N) atoms, enabling only identification of Si lattice sites. In contrast, under the annular bright-field (ABF) STEM imaging, Si atoms reduce the electron intensity reaching the ABF detector by scattering more electrons away from the ABF detector's collection region, thus appearing darker and enabling simultaneous visualization of both Si and C(N) atoms (Extended Data Fig. 4a)[32]. From the selected area electron diffraction (SAED) pattern (Fig. 2b), the interplanar spacings $d_{(\bar{1}1\bar{1})}$ and $d_{(11\bar{1})}$ are 0.253 nm and 0.250 nm, and the angles between the $(00\bar{2})$ and $(\bar{1}1\bar{1})$ planes and between the $(00\bar{2})$ and $(11\bar{1})$ planes are 55.2° and 54.3°, respectively. In contrast, for the epitaxial 3C-SiC with low carrier concentration ($n_e$) of ~$8.6 \times 10^{15}$ cm$^{-3}$, the $d_{(\bar{1}1\bar{1})}$ and $d_{(11\bar{1})}$ values are 0.254 nm and 0.252 nm, as determined from the fast Fourier transform (FFT) pattern (Fig. 2c and Extended Data Fig. 4b), and the angles between the $(00\bar{2})$ and $(\bar{1}1\bar{1})$ planes and between the $(00\bar{2})$ and $(11\bar{1})$ planes are 54.7° and 54.6°, respectively. These results demonstrate an *R3m* lattice distortion along the parent cubic [111] direction in the heavily N-doped 3C-SiC single crystals, whereas the lightly N-doped epitaxial films retain their cubic structure, further validating the PJTE-induced lattice distortion[26,33].

XRD phi-scans demonstrate a symmetry breaking from the pristine four-fold and three-fold azimuthal symmetries of the {111}and {200} reflections (Fig. 2d,f and Extended Data Fig. 4c,d). This departure, along with the splitting of the (111) peak (Fig. 2e), provides macroscopic evidence of the lattice distortion[27,33]. Clear second harmonic generation (SHG) peaks at 400 nm appear upon excitation by the 800 nm laser for heavily N-doped 3C-SiC single crystals (Fig. 2g and Extended Data Fig. 5a). A log-log plot of the SHG intensity versus the excited laser power reveals a linear relationship with a slope of ~ 2.01 (Fig. 2h), which is consistent with the fundamental relation of $I_{SHG} \propto P_{Laser}^2$ and thus confirms that the SHG is an intrinsic property[34-37]. Polarization-resolved azimuthal SHG exhibits a well-defined six-fold symmetric pattern (Fig. 2i and Extended Data Fig. 5b), revealing the $R3m$ lattice distortion according to: $I(\theta) \propto \cos^2 3\theta \propto \frac{1+\cos 6\theta}{2}$[34,37]. However, no SHG signal is observed in the epitaxial film (Extended Data Fig. 5c), further showing the carrier induced structure distortion. The $R3m$ phase becomes thermodynamically favorable at moderate electron doping, with ~$2.9 \times 10^{20}$ cm$^{-3}$ located in the critical doping range (Extended Data Fig. 6a). The crystallographic parameters for the $R3m$ phase are: $a = b = 3.0849(1)$ Å, $c = 7.552(4)$ Å, $\alpha = \beta = 90°$, $\gamma = 120°$ (hexagonal setting), determined from the single crystal XRD results (Supplementary Table 2). The $[0001]_R$ polar axis of $R3m$ corresponds to the $[111]_C$ direction of the parent cubic $F\bar{4}3m$ lattice[33].

**Switchable ferroelectric polarization under external voltages**

The heavily N-doped 3C-SiC single-crystal wafers were characterized by piezoresponse force microscopy (PFM). Obvious contrast is seen in the PFM phase and corresponding PFM amplitude images after a box-in-box poling under a tip bias of ±7 V (Fig. 3a, Supplementary Fig. 1), verifying the remanent polarization and thus the bistable switchable polarization, although no 180° phase contrast is obtained in the images owing to the screening effects from the free carriers in the ferroelectric metals[39,40]. Nevertheless, 180° polarization reversal is clearly seen in the local PFM phase loops (Fig. 3b), consistent with prior observations reported for other metallic ferroelectrics[6]. Well-defined butterfly amplitude loops (Fig. 3c) further confirm the switchable polarization. To rule out charge injection or the artifact-induced PFM phase and amplitude change, the local PFM phase (Fig. 3d) and amplitude (Fig. 3e) loops were recorded at an ac frequency of 0.1-3 Hz. The hysteresis and amplitude loops remain robust rather than disappearing as the frequency increases, confirming the intrinsic polarization reversal rather than charge injection or other artifacts[38,40]. The coercive voltage is about 0.4 V (Fig. 3b), which is lower than those (~1-50 V) of $BaTiO_3$ single crystals and comparable to those of ultrathin ferroelectric films (Fig. 3f and Supplementary Table 3). The low coercive voltage is compatible with lower power operation, laying a solid material foundation for the fabrication of low-voltage, non-volatile, ferroelectric devices.

Although the switchable ferroelectric signals are detected by PFM, this technique alone has limits in providing definitive evidence for intrinsic FE, particularly for a metallic system. To directly probe the electric-field-driven polarization reversal at the atomic scale, we performed in-situ TEM and STEM measurements under external voltages of +4 V and −4 V (Fig. 3g-l, Supplementary Fig. 2-3).

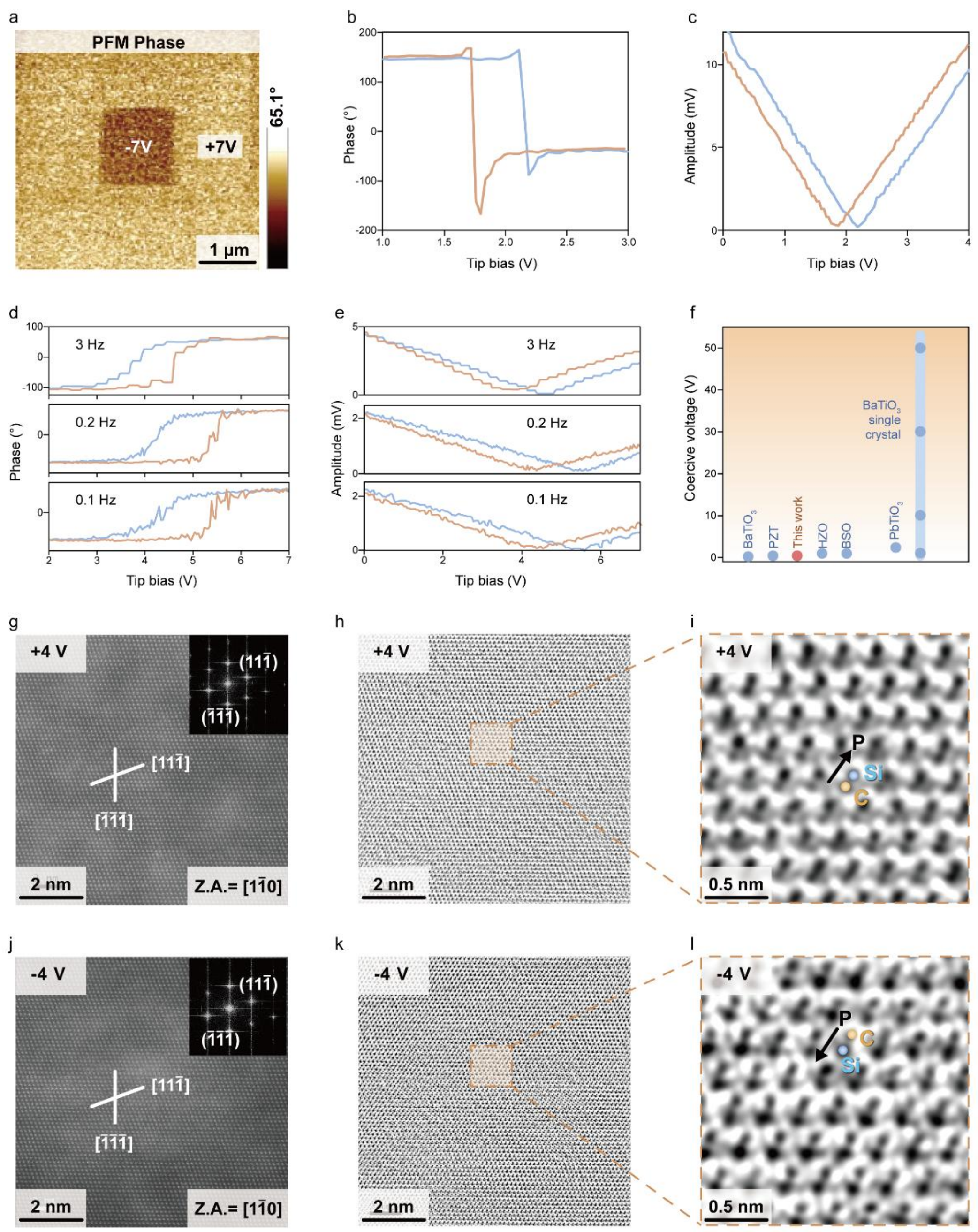

**Fig. 3 | Ferroelectric behaviors of heavily N-doped 3C-SiC verified by PFM and in-situ STEM. a,** A PFM phase image after box-in-box writing with +7 V for 3×3 μm, and −7 V for 1×1 μm regions by applying a tip bias. Well-defined PFM phase and amplitude contrast is obviously seen, confirming the remanent polarization states and thus switchable polarization. **b,** Local PFM phase loops and corresponding (**c**) PFM amplitude butterfly loops acquired on the poled regions. The results further confirm switchable polarization. **d,** Frequency-dependent PFM phase loops (0.1-3 Hz) and (**e**) corresponding PFM amplitude curves, showing the frequency independence of the PFM phase loops and PFM amplitude butterfly loops. **f,**

Comparison of coercive voltage with other ferroelectric materials. HAADF-STEM images viewed along the $[1\bar{1}0]$ zone axis acquired at biases of (**g**) +4 V and (**j**) −4 V. The insets of (**g**) and (**j**) show the corresponding FFT patterns. The corresponding ABF-STEM images measured at biases of (**h**) +4 V and (**k**) −4 V. **i,l,** The magnified ABF-STEM images of the regions indicated by the filled rectangles in panels (**h**) and (**k**), respectively. From the ABF-STEM results, a polarization reversal of approximately 180° along the <111> direction under an external voltage is directly visualized at the atomic scale. Z.A., zone axis.

A bright-field TEM image acquired from the 3C-SiC (111) reflection under a two-beam condition, with the specimen tilted approximately 5° away from the [110] zone axis, shows a clear contrast change under +4 V and -4 V bias (Supplementary Movie 1). This contrast evolution supports a reversible voltage-induced ferroelectric domain switch or structural change associated with polarization reversal[41,42]. The corresponding SAED patterns show only minor changes in the *d*-spacings under opposite biases (Extended Data Fig. 7), suggesting that no significant lattice expansion or contraction occurs during switching. Thus, the contrast change in the bright-field TEM image under opposite biases is due to a ferroelectric domain change. The HAADF-STEM images (Fig. 3g,j) also show an unchanged crystal lattice under external voltage bias. The ABF-STEM images reveal a reversible change in the projected polar displacement between the Si and C/N atomic columns. Under +4 V, the polar displacement is oriented along the [111] direction (Fig. 3h,i, Extended Data Fig. 7), consistent with the PFM response. When the bias is reversed to −4 V, the projected polar displacement switches toward the opposite $[\bar{1}\bar{1}\bar{1}]$ direction (Fig. 3k,l, Extended Data Fig. 7), indicating an approximately 180° reversal of the local polar distortion. These observations provide atomic-scale evidence for voltage-induced switching of the polar distortion in the metallic ferroelectric-like state. The reversible modulation of the Si-C/N relative displacement along the <111> direction may contribute to the observed FE response.

First-principles calculations reveal a symmetric double-well profile for the down and up polarization along the *c*-axis of the *R*3*m* structure at the minimum energy (Extended Data Fig. 6b), a characteristic feature of ferroelectric materials. Thus, both theoretical and experimental results verify the existence of a switchable polar order.

The Curie temperature is up to at least 599 °C determined from the Raman results (Fig. 1i). The Curie temperature is expected to exceed the operating temperature of an $Al_{0.68}Sc_{0.32}N$ device (~600 °C)[28,29], surpassing the ferroelectric transition temperature of conventional FE materials such as $BaTiO_3$ (~120 °C)[30], $PbTiO_3$ (~490 °C)[15], PZT (230-490 °C)[15], and representative polar (ferroelectric) metals or superlattice such as $LiOsO_3$ (-133 °C)[3], $(SnSe)_{1.16}(NbSe_2)$ superlattice (~110 °C)[6], $WTe_2$ (~67 °C)[5], see Supplementary Table 3.

The observation of robust FE in this metallic system essentially arises from the incomplete screening of local polarization fields by free carriers[9,43]. The PDOS results show that the N-derived orbitals at the Fermi level are spatially decoupled from the hybridized N-Si-C bonds that drive polar distortion, as evidenced by electron localization function (ELF) analysis (Extended Data Fig. 6c-d) and the calculated PCD (left inset of Fig. 1e). Confined to antibonding regions and unable to cross the nodal planes between atoms, conduction electrons in antibonding states cannot fully screen the bond-axis electric field. This leaves the lattice polar and ferroelectric-like, while the electrons still conduct by hopping parallel to the nodal planes. The Thomas-Fermi screening length ($\lambda_{TF}$) is calculated to be 1.4 nm using the three-dimensional free electron gas model with $n = 2.9\times10^{20}$ cm$^{-3}$ and $\varepsilon_r = 9.7$ (Supplementary text)[9,43], which is larger than the lattice constant ($c$ = 0. 7552(4) nm) and Si-C bond length (~0.189 nm). This result further confirms the incomplete screening of local bond dipoles by free carriers. However, this estimation should be viewed as a rough approximation, as the Thomas-Fermi model assumes a homogeneous and isotropic electron gas. Therefore, this approximation likely underestimates the effective screening length along the polar direction, as it neglects the spatial confinement of antibonding electrons to nodal-plane regions. The actual screening of the bond-axis dipoles is expected to be even weaker.

**Ferroelectric tunnel junction devices**

Ferroelectric tunnel junctions (FTJs) have wide applications in binary data storage and memristors with advantages of non-destructive readout and simpler device architecture[44-52]. We fabricated and characterized wafer-scale FTJs arrays with a structure of Au/3C-SiC/Au (Fig. 4a). Distinct HRS and LRS are observed in the I-V curve at an operating bias of 1 V (Extended Data Fig. 8a-c,h, Supplementary Fig. 4), indicating an ultralow switching voltage of 1 V. HRS and LRS switching is further confirmed under biases of 2-4.5 V (Fig. 4b,c, Extended Data Fig. 8d-f, Supplementary Fig. 4). The observation of stable, bipolar, non-volatile switching distinctly reveals the switchable ferroelectric polarization, which modifies the Schottky barrier height at the Au/3C-SiC interface upon polarization reversal[44]. The on/off ratio increases with the applied bias, reaching ~20 at 4 V (Extended Data Fig. 8g). High carrier concentration, which partially screens the polarization-induced barrier modulation, limits the on/off ratio. Benefiting from the metallicity, the devices exhibit an ultrafast response speed of ~50 ns (a delay time ($t_d$) of ~15 ns and a rise time ($t_r$) of ~35 ns) (Fig. 4d,e). The response speed is among the fastest reported values for ferroelectric-based memories[47].

The HRS and LRS resistances were measured after different cycling and retention times (Fig. 4f, Extended Data Fig. 9a). It is seen that the resistances of the HRS and the LRS are quite stable even after operation for $10^4$ s and 8873 cycles (Fig. 4f). Notably, the HRS and LRS resistances remain well-separated over $10^5$ s and 85927 cycles (Extended Data Fig. 9a). A projected retention time reaches 100 years based on short-term measurements (Extended Data

Fig. 9b). A comprehensive benchmarking against state-of-the-art FTJs in terms of switching speed, endurance, and retention (Supplementary Table 4) shows the superior performance of our devices with a response time of ~ 50 ns, an endurance exceeding 85927 cycles, and a projected retention time of 100 years. By combining wafer-scale dimensions, bulk single crystal, low resistivity, and switchable ferroelectric polarization, the N-doped 3C-SiC presented here simultaneously offers an ultrafast write speed, an ultralow operating voltage, and excellent anti-fatigue and endurance characteristics, outperforming previously reported ferroelectric/polar metals and most FTJs (Extended Data Fig. 8i and Supplementary Tables 4,5).

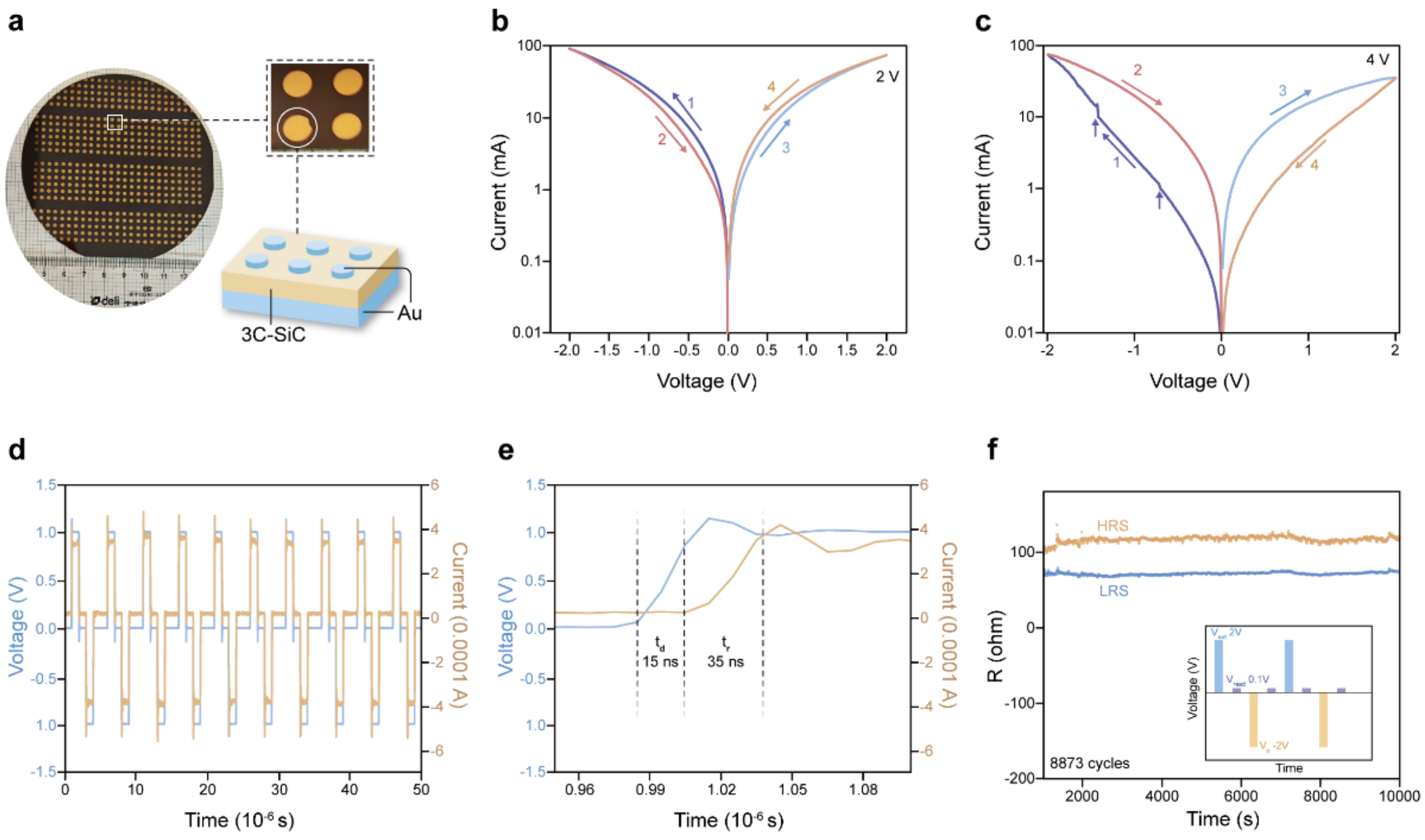

**Fig. 4 | Measured performance of heavily N-doped 3C-SiC devices. a,** Photographs of the FTJ array fabricated on a 4-inch heavily N-doped 3C-SiC single crystal. Insets: magnified views of individual FTJs (top left) and a schematic cross-section of the device structure (top right). **b,c,** Evolution of the I-V hysteresis loops on a log-log scale under positive pre-poling voltages ranging from 2 V to 4 V. Increasing the pre-poling voltage promotes saturated upward polarization and enables bidirectional reversible domain switching. The HRS and LRS are clearly distinguishable at a read voltage of 1 V. **d,** Ultrafast response with excellent reproducibility and consistency over 10 cycles. **e,** Magnified image of **(d)** at the time range of 950-1100 ns, indicating an ultrafast response time of ~50 ns. **f,** Resistance measurements under +2 V/-2 V voltages pulses over 8873 cycles and 10000 s. Inset of (**f**): The pulse sequence used during the endurance and retention tests. Each cycle consists of a +2 V pulse, a 0.1 V read, a -2 V pulse, and a 0.1 V read.

## Summary and Outlook

In summary, we break the mutual exclusivity between metallicity and ferroelectricity in a non-polar covalent semiconductor via orbital selective doping. This strategy can be readily applied to other systems like silicon, diamond, and so on. Our work opens several exciting avenues for future research. First, it establishes doping-induced ferroelectricity in non-polar semiconductors as a viable design strategy. Second, the coexistence of free carriers and polar order invites investigation of emergent phenomena, including the interplay between ferroelectricity, electron-phonon coupling, and polaron dynamics. Looking forward, further studies on the formation dynamics of ferroelectric domains, their atomic-scale spatial distribution, and the dynamic process under external voltage will be essential to fully understand this class of materials and harness their full potential for next-generation ferroelectric devices.

**Acknowledgements** This work is supported by the Beijing Natural Science Foundation (nos. F261026, Z250020), National Key Research and Development Program of China (no. 2024YFE0205400), National Natural Science Foundation of China (nos. 52572017, 52525207, and 52288102), and IOP, CAS (E1K2161MA2). Thanks for Q. Zheng, H. Gao, Y. Gong for the discussion for the TEM measurements.

**Author contributions**. X. C. and H. L. designed and supervised this project; H. L., Y. Y., G. W. and Z. Z. grew the single crystals. J. H., A. N., Y. F., Y. T., and H. L. designed and performed the TEM characterization; H. L., S. D., and Y. X. conducted SHG measurements. H. L. and Y. Y. conducted APT and Hall measurements; H. L., Z. S., D. G. conducted XRD measurements; H. L., Y. F. conducted PFM and *I-V* measurements; Q. W. and H. P. designed and conducted the devices measurements. Z. L. performed the first-principles calculations; J. Y., H. S. performed the epitaxial growth; J. G. and W. W. supported the project; All authors participated in discussions of the research.

**Competing interests.** The authors declare no competing interests.

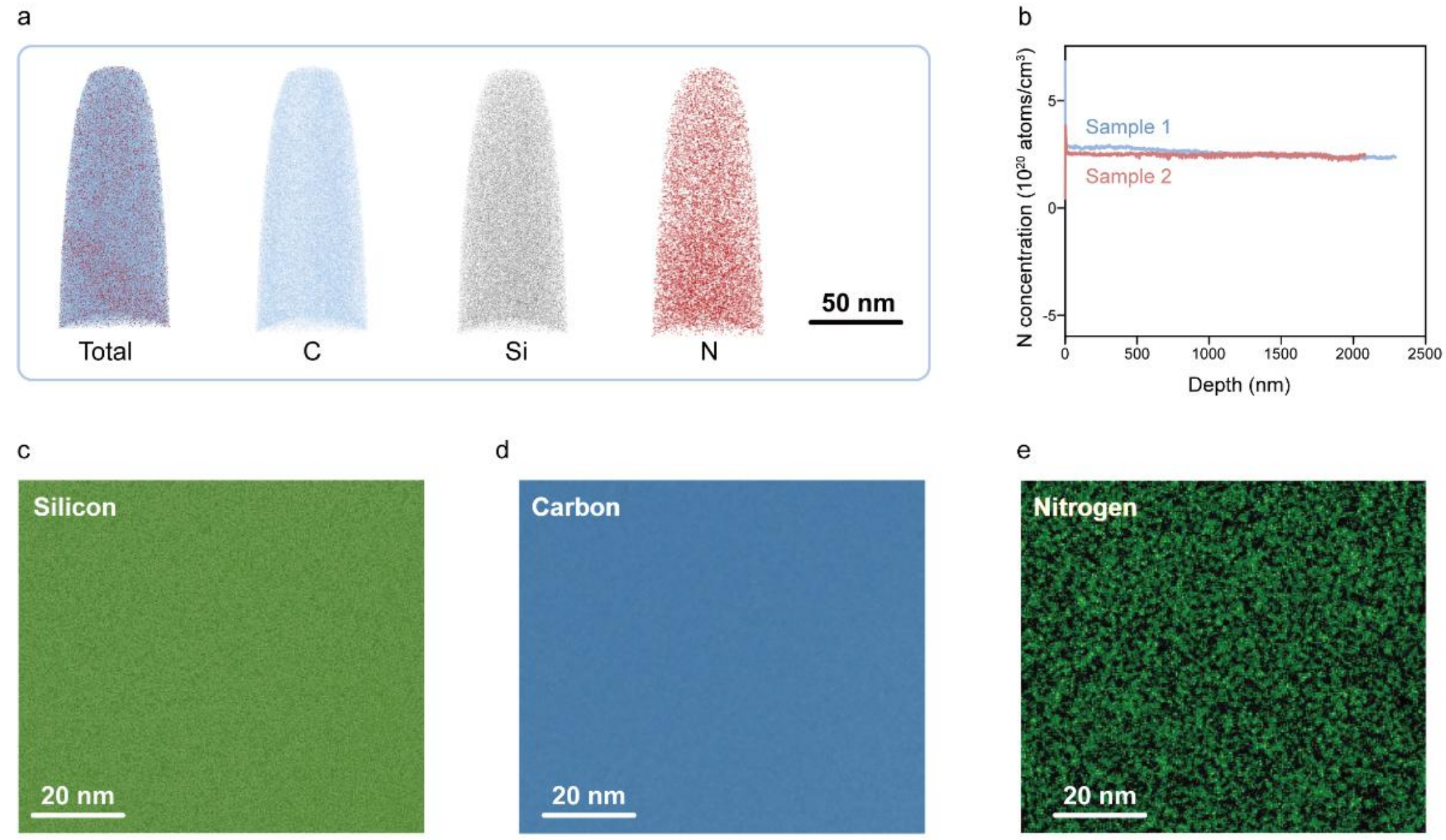

**Extended Data Fig. 1 | Elemental distribution within the heavily N-doped 3C-SiC single crystal. a,** APT elemental maps (Total, C, Si, and N) show the uniform distribution of N in the lattice. **b,** Depth-dependent N concentration profile measured by SIMS in two samples, showing a uniform doping level of ~$2.9 \times 10^{20}$ atoms·$cm^{-3}$ over a depth of 2.0 μm from the surface. EELS mapping results for (**c**) Si, (**d**) C, and (**e**) N also confirm the uniform N distribution in the sample.

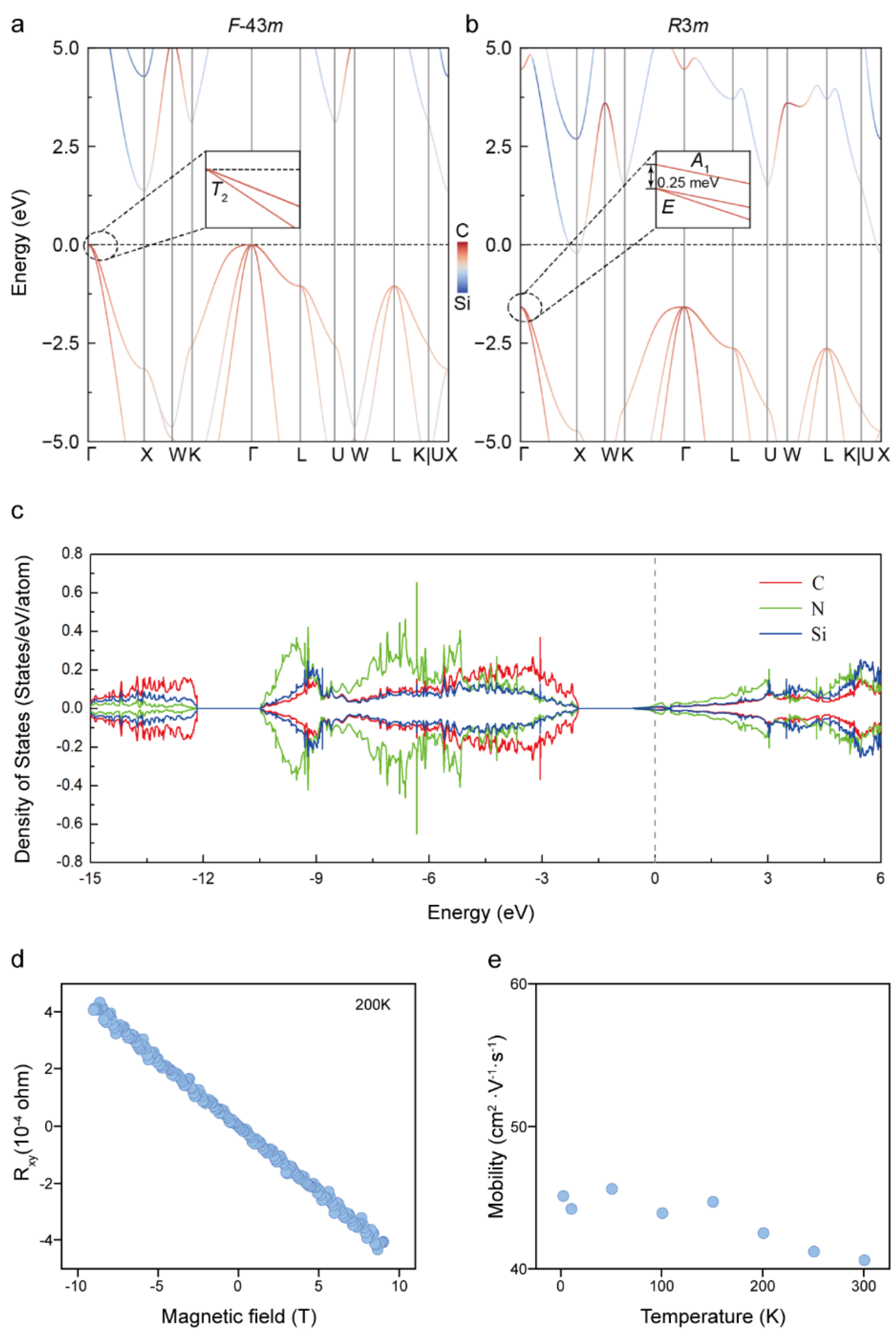

**Extended Data Fig. 2 | Electronic structure difference between the SiC with $F\bar{4}3m$ and $R3m$ phases from first-principles calculations with a carrier concentration of ~$2.9\times10^{20}$ $cm^{-3}$.** Electronic structures for (**a**) SiC with a space group of $F\bar{4}3m$ and (**b**) SiC with a space group of $R3m$. The insets show enlarged views of the valence-band maximum (VBM) labeling the corresponding irreducible representations. **c,** Calculated PDOS results for N-doped 3C-SiC, verifying non-zero states at the Fermi level (0 eV, dashed line). This result shows the metallicity and strong orbital hybridization among N, Si, and C that induces the structural phase transition. **d,** Hall resistance measured at 200 K. **e,** Carrier mobility in the temperature range of 2−300 K.

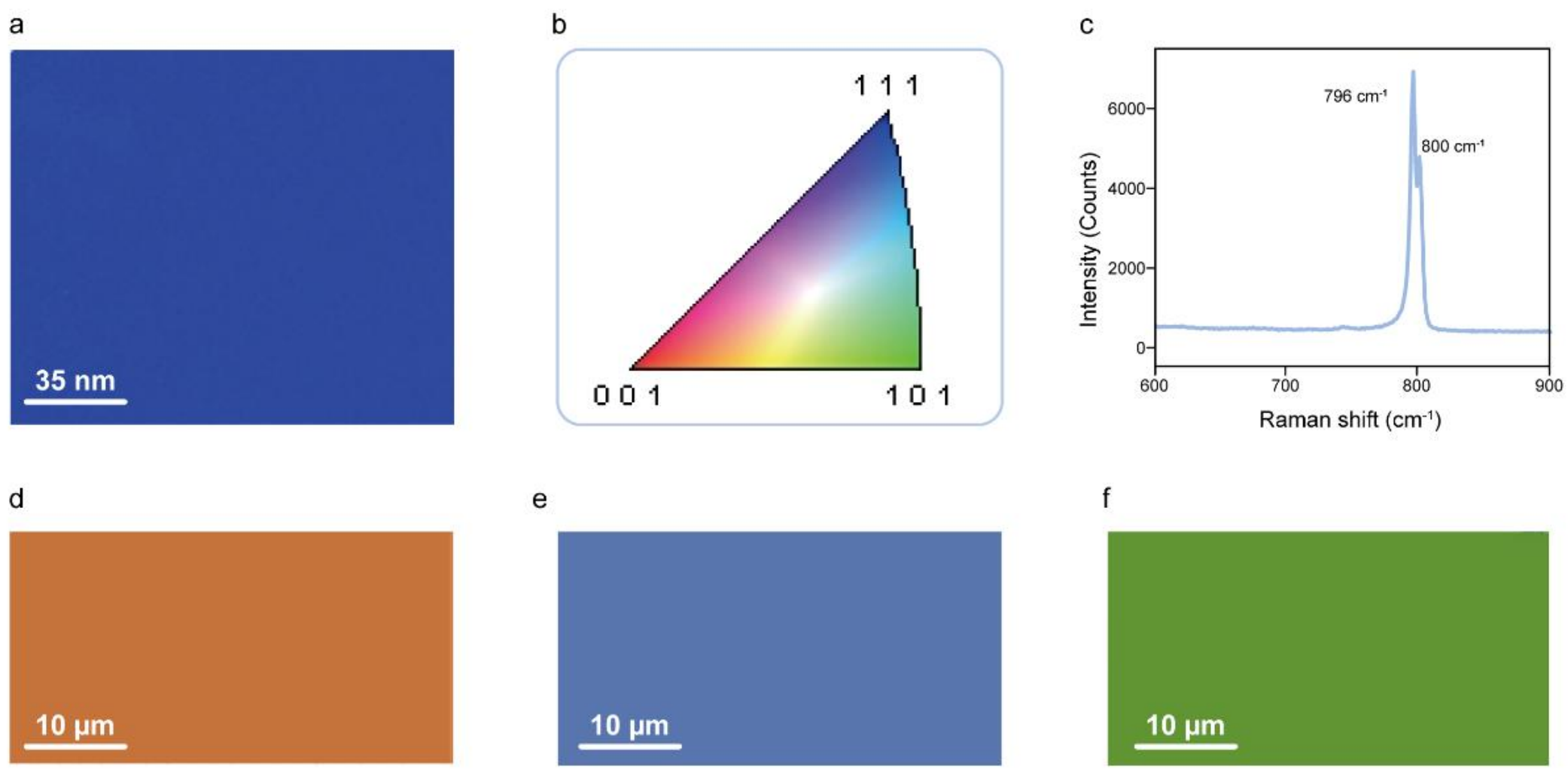

**Extended Data Fig. 3 | EBSD and Raman characterizations of 6-inch heavily N-doped 3C-SiC single crystals.** (**a, b**) EBSD results show the (111) growth front for N-doped 3C-SiC single crystals. (**c**) TO Raman peaks are split into two peaks located at 796 $cm^{-1}$ and 800 $cm^{-1}$ measured on the (111) plane. (**d-f**) Raman mapping images for the 796 $cm^{-1}$ peak measured on three random regions of a 6-inch wafer as shown in the inset of (**c**). The mapping results confirm the single-phase nature of the heavily N-doped 3C-SiC single crystals.

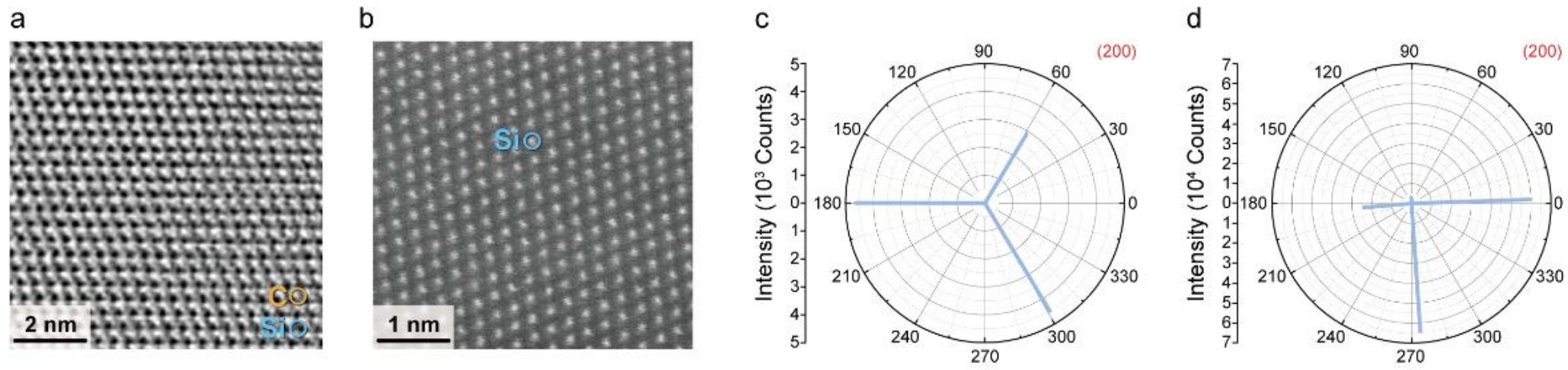

**Extended Data Fig. 4 | Structural characterizations of heavily N-doped 3C-SiC and lightly N-doped epitaxial 3C-SiC film with a carrier concentration of ~8.6×$10^{15}$ $cm^{-3}$. a,** ABF-STEM image of heavily N-doped 3C-SiC single crystals. The Si and C(N) atoms are clearly visible in the ABF-STEM image viewed along the [1$\bar{1}$0] zone axis. **b,** Atomic-resolution HAADF-STEM image of the epitaxial 3C-SiC film viewed along the [1$\bar{1}$0] zone axis. **c,d,** XRD $\varphi$-scans of the (200) and (111) plane with out-of-plane orientations along [111] and [100] direction in polar modes, respectively. The deviation of the three-fold and four-fold azimuthal symmetries of the (200) and (111) reflections is clearly observed, indicating the lattice distortion of the heavily N-doped 3C-SiC single crystals.

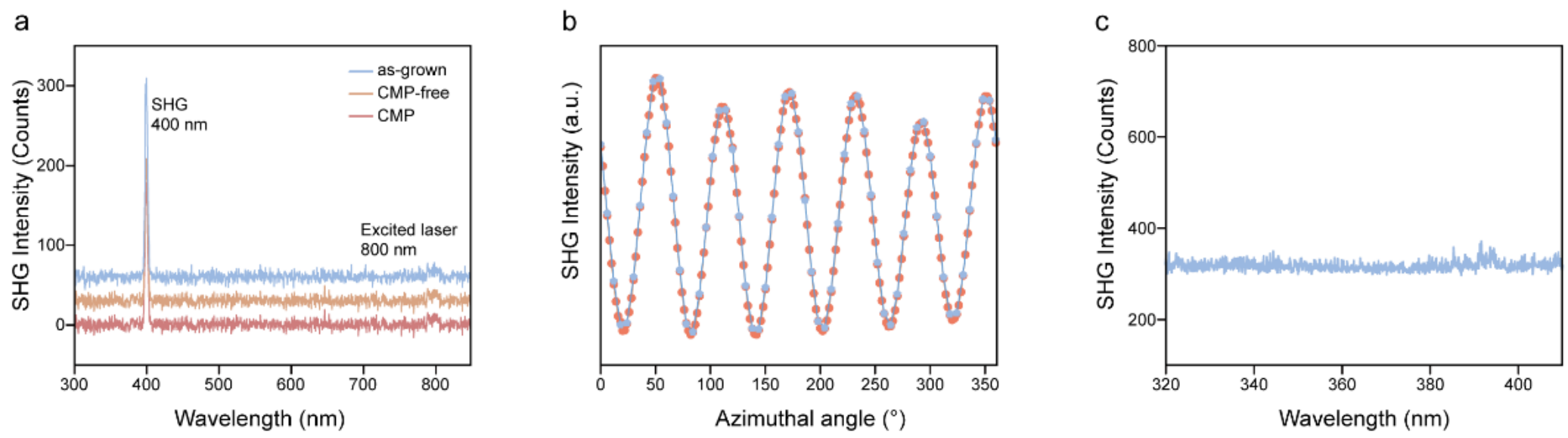

**Extended Data Fig. 5 | SHG characterizations of heavily N-doped 3C-SiC single crystals and low N-doped epitaxial 3C-SiC film with a carrier concentration of ~8.6×10$^{15}$ cm$^{-3}$. a,** SHG spectra (excited with an 800 nm laser) of as-grown, CMP-treated, and CMP-free N-doped 3C-SiC single crystal wafers. **b,** Azimuthal angle dependence of the SHG spectra measured at two random regions of CMP-treated N-doped 3C-SiC single crystals, excited with a 780 nm laser. **c,** SHG spectra excited with a 780 nm laser for the epitaxial 3C-SiC film, revealing the absence of an SHG signal.

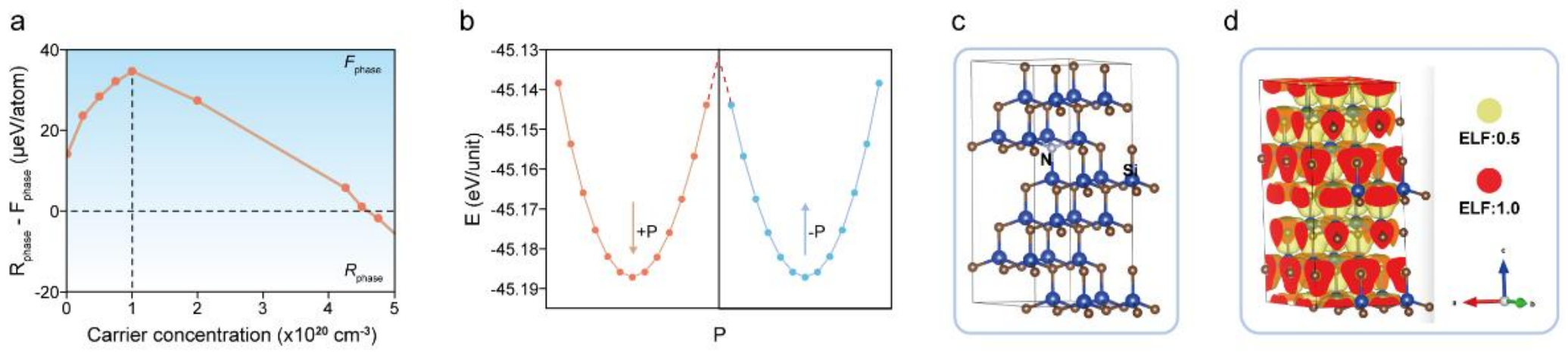

**Extended Data Fig. 6 | First-principles calculation results for heavily N-doped 3C-SiC. a,** Energy difference between the $R3m$ phase and the $F\bar{4}3m$ phase as a function of carrier concentration at 0 K. Positive values correspond to the thermodynamic stability of the $F\bar{4}3m$ phase, whereas negative values correspond to the dominant $R3m$ distorted phase. **b,** Calculated total energy versus ferroelectric polarization for two opposite polarization directions along $[00\bar{1}]$ and [001] direction according to $R3m$ phase. The red and green data curves correspond to downward and upward spontaneous polarization, respectively, showing a symmetric double-well structure for +P and −P at the minimum energy. **c,** Crystal structure and **d,** corresponding ELF distribution of heavily N-doped 3C-SiC based on the $R3m$ phase. ELF iso-surfaces clearly distinguish localized electrons in Si-C/Si-N covalent bonds (the origin of ferroelectric polarization) from delocalized free-electron gas (the origin of metallic conduction), revealing the spatially separated mechanism underlying their coexistence.

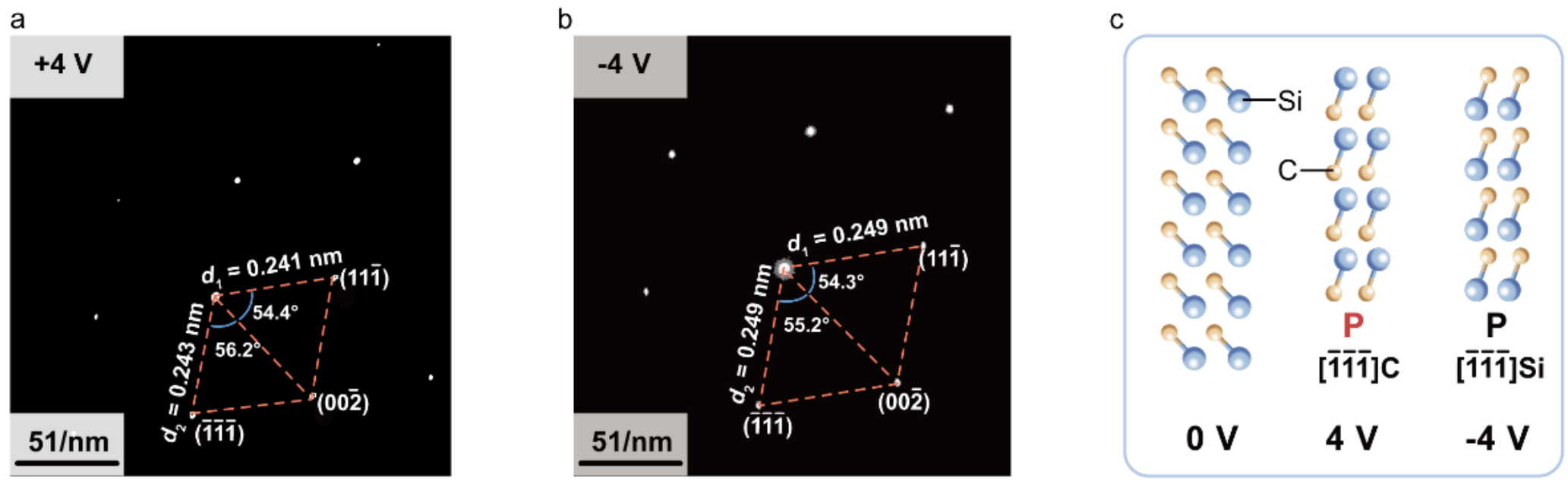

**Extended Data Fig. 7 |** SAED patterns for the heavily N-doped 3C-SiC lattice viewed along the [1$\bar{1}$0] zone axis under biases of (**a**) +4 V and (**b**) −4 V. **c,** Illustration of the polarization direction under opposite biases.

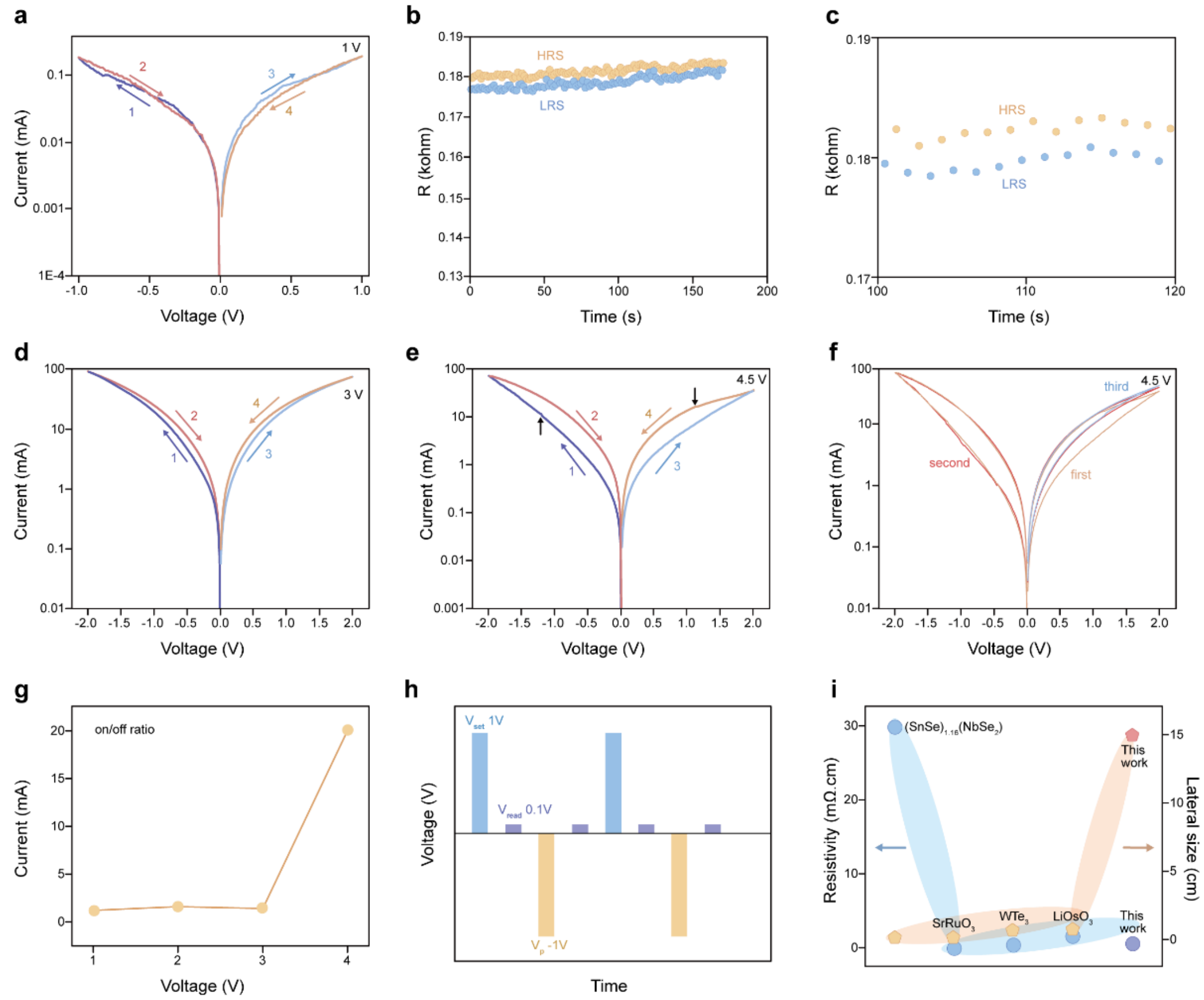

**Extended Data Fig. 8 | Performances of heavily N-doped 3C-SiC based FTJs. a,** I-V hysteresis loops on a log scale after a 1 V bias. **b,c,** Resistance switching between the on and off state measured at 0.1 V after the application of 1 V and -1 V voltage pulses over 100 cycles. **d-e,** Evolution of I-V hysteresis loops on a log scale under positive pre-poling voltages of 3 V to 4.5 V. **f,** Reproducible I-V hysteresis curves over three consecutive cycles, verifying the robust ferroelectric switching stability. **g,** On/off ratio under different biases. **h,** Illustrations of the applied voltage pulse sequences for the anti-fatigue measurements: Vset and Vp of 1 V and -1 V. **i,** Benchmark comparison plot of lateral size versus room-temperature resistivity with state-of-the-art ferroelectric/polar metals, highlighting the unique advantage of our work: combining wafer-scale dimension and low metallic resistivity.

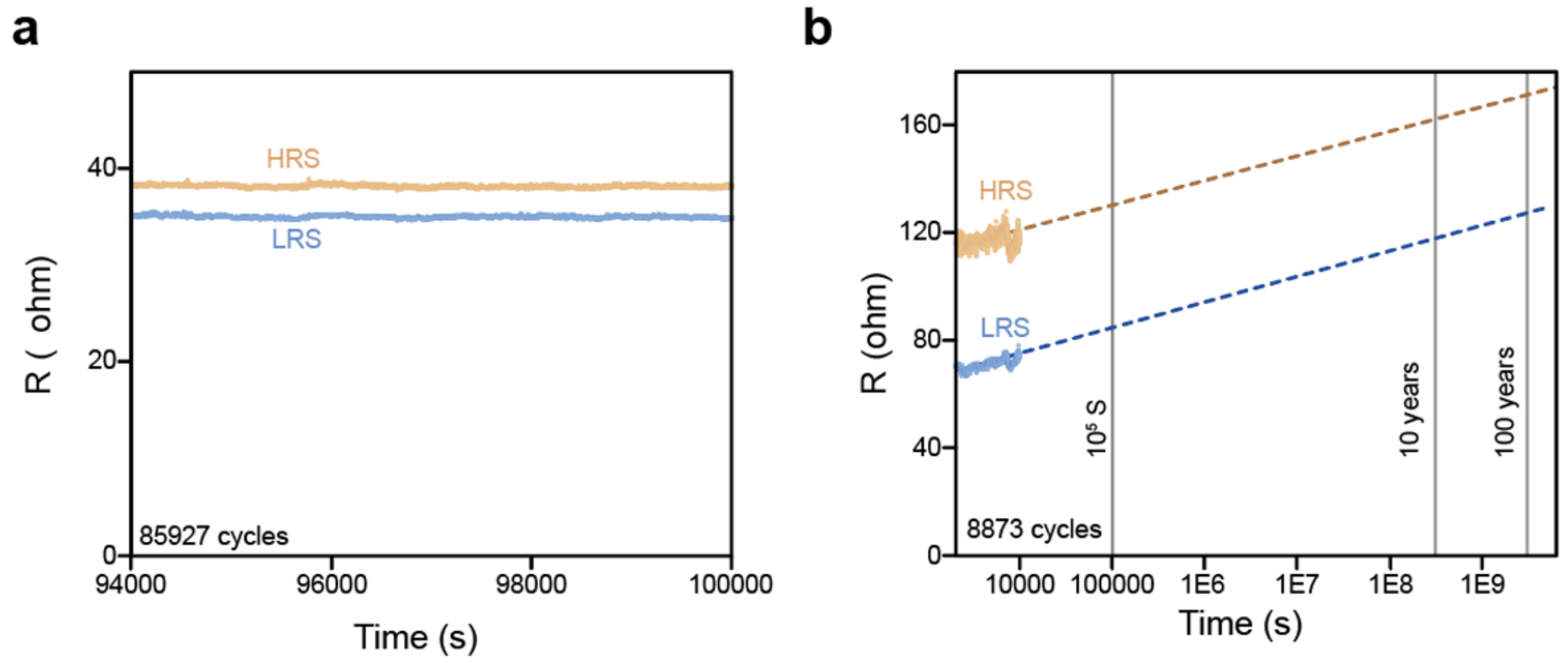

**Extended Data Fig. 9 | Endurance and retention results for FTJs. a,** Resistances of HRS and LRS over $10^5$ s and 85927 cycles. **b,** The projected retention time reaches 100 years based on short-term measurements. Dashed lines represent extrapolation.

Supplementary Information

# Breaking the mutual exclusivity between metallicity and ferroelectricity in a non-polar covalent semiconductor via orbital selective doping

Hui Li,[1] Yunfan Yang,[1,8] Junquan Huang,[2] Yukun Feng,[1] Guobin Wang,[3] Qinci Wu,[4] Jun Deng,[5] Zhaolong Liu,[1,8] Subi Du,[1,8] Dongliang Gong,[6] Zaihui Shen,[1,8] Anmin Nie,[2*] Yang Xu,[1,8] Junwei Yang,[7] Zesheng Zhang,[3] Huaping Song,[7] Jiangang Guo,[1,8] Wenjun Wang,[1,8] Hailin Peng,[4] Yongjun Tian,[2] and Xiaolong Chen*[1,8]

[1]Beijing National Laboratory for Condensed Matter Physics, Institute of Physics, Chinese Academy of Sciences, Beijing 100190, China

[2]Center for High Pressure Science, State Key Laboratory of Metastable Materials Science and Technology, Yanshan University, Qinhuangdao 066004, China

[3]Beijing Lattice Semiconductor Co., Ltd., Beijing 101300, China

[4]Center for Nanochemistry, Beijing Science and Engineering Center for Nanocarbons, Beijing National Laboratory for Molecular Sciences, College of Chemistry and Molecular Engineering, Peking University, Beijing, P. R. China.

[5]Center for High Pressure Science and Technology Advanced Research, Beijing, 100193, China.

[6]Institute of Electrical Engineering, Chinese Academy of Sciences, Beijing 100190, China

[7]PowerEpi Semiconductor Co., Ltd. Dongguan, Guangdong 523808, China

[8]School of Physical Sciences, University of Chinese Academy of Sciences, Beijing 100049, China

Corresponding authors: Anmin Nie, E-mail: anmin@ysu.edu.cn; Xiaolong Chen, E-mail: chenx29@iphy.ac.cn

**Materials and Methods**

The N-doped 3C-SiC bulk single crystals with a diameter of 6-inch were grown from high-temperature solutions, as reported before[23,24]. The main difference for the 6-inch 3C-SiC is the size of the seed crystal and the graphite crucibles. For the growth of 6-inch N-doped 3C-SiC single crystals, a semi-insulating 4H-SiC with a zero-degree (0001) Si-terminated surface was applied as the seed crystal. After the growth, the single crystal wafers with a thickness of 350-micron were cut from the as-grown boules and undergone grinding, polishing, and chemical mechanical polishing (CMP) processes. The samples cut from the 3C-SiC single crystal wafers were then used to conduct different measurements. The low N-doped 3C-SiC epitaxial films were grown via a chemical vapor deposition method, as reported before[23]. The samples cut from the epitaxial films were used to conduct the following measurements.

Vertical FTJs were fabricated by depositing 50 nm Au electrodes via thermal evaporation on both sides of the sample using home-made equipment, with a mask applied to one side. The *I-V* curves were collected using a Keithley 2400 via the self-written software. The samples were pre-biased using 1.0-4.5 V. The *I-V* curves were measured from -2.0 V to 2.0 V with a step of 0.01 V. For long-term retention of polarization-modulated states, ± (1.0, 2.0, 3.0, 4.0) V pulsed voltages were applied at different intervals ranging from ns to s with a 0.1 V read voltage. The on/off ratio is the maximum ratio between the current at the low-resistive state (LRS) and that at the high-resistive state (HRS) at the same read voltage[53].

For APT measurements, a needle-shaped sample was obtained by focused ion beam (FIB) milling with a FEI Helios G5 Dual Beam system. After that, a 50-nm Pt layer was deposited on the sample. The measurements were performed by a LEAP 5000 XR from Cameca with a voltage between 5 and 9 kV with 355-nm laser pulses (30 pJ, 200 kHz) for sputtering the sample at 40 K and $1\times10^{-11}$ torr. The three-dimensional (3D) distribution of the elements Si, C, and N was reconstructed using Cameca AP Suite 6.3 software.

Dynamic SIMS (D-SIMS) analysis utilizing a cesium ($Cs^+$) primary ion beam at a 9 keV impact energy in negative secondary ion mode was performed on a Cameca 7f Auto DSIMS system.

EBSD measurements were performed in the in-plane direction with a step size of 1 μm. A Hitachi SU5000 scanning electron microscope equipped with an EDAX Hikari EBSD detector was used for data acquisition. The measurement procedure included sample surface preparation, selection of the region of interest under SEM, setting of the EBSD scan parameters, collection of orientation data, and subsequent analysis of grain orientation, inverse pole figure (IPF) maps, and related EBSD data.

X-ray diffraction (XRD) was performed using a Bruker D8 Discover diffractometer (Cu Kα, $\lambda$ = 1.5406 Å). The XRD phi scans were conducted on (111) and (100) growth front surfaces at room temperature using Empyrean X-ray equipment at 45 kV and 40 mA. The step size for the scan was 0.01°, the time per step was 0.2 s, and the scan speed was 0.01°/s. XRD φ-scan profiles for the (111) plane (out-of-plane [100]) and the (200) plane (out-of-plane [111]) reveal deviations from the four-fold and three-fold azimuthal symmetries of the {111} and {200} reflections, respectively[54,55]. Single-crystal X-ray diffraction (SC-XRD) measurements were performed on a Bruker D8 Venture diffractometer with a Mo $K\alpha$ radiation source ($\lambda$ = 0.71073 Å) at room temperature (298 K). The SC-XRD data were processed using APEX software.

Raman spectra were collected on a Horiba LabRAM HR Evolution (532 nm excitation, 10 mW) from -193 ºC to 599 ºC with spatial and spectral resolutions of 1 μm and 0.5-1.0 $cm^{-1}$, respectively. Raman spectra and Raman mapping were acquired on a Horiba XploRA Plus (532 nm excitation, 10 mW) with an acquisition time of 30 s per spectrum under three accumulations.

Second-harmonic generation (SHG) measurements were conducted on a home-made SHG system with a femtosecond laser (800 nm, 100 fs, 80 MHz) with the excitation laser perpendicular to the sample surface. For angle-dependent SHG measurements, the sample was rotated with the laser perpendicular to the sample on the home-made system with a central laser wavelength of 780 nm (with a pulse duration of ~100 fs and a repetition rate of 100 MHz). The root mean square roughness for the CMP N-doped 3C-SiC single crystal wafers is ≤ 0.2 nm.

PFM was conducted on an atomic force microscope (Multimode 8, Bruker) using conductive SCM-PIT-V2 tips (antimony-doped Si covered by Pt/Ir). For PFM measurements, the 5×5 μm region was pre-poled by applying a −5 V DC bias to the conductive tip. Then, a +7 V voltage was applied to the conductive tip to polarize a 3×3 μm area. After that, a −7 V voltage was applied to the conductive tip to polarize a 1×1 μm area. For scanning, the DC voltage was set to zero while an AC voltage of 1 V at 1 Hz was applied to the conductive tip. To verify the intrinsic ferroelectric response of the sample, PFM phase and amplitude hysteresis curves were recorded with AC driving frequencies ranging from 0.1 to 3 Hz[40, 56].

For in-situ TEM and STEM measurements, heavily N-doped 3C-SiC single crystal samples were prepared by a standard FIB lift-off process (FEI SCIOS 2 using $Ga^+$ ions) onto a P.J.B.DT.2 chip produced by DENS[57]. Before mechanical cutting, a 1.5 μm Pt layer was deposited for sample protection during the slicing process. During the FIB cutting process, 30 kV and 5 nA were used for coarse milling, followed by 30 kV and 3 nA for fine milling. For thinning the sample, the following parameters were applied: 30 kV and 0.1 nA, 16 kV and 50 pA, and 8 kV and 8.9 pA. To remove the amorphous layer and avoid the $Ga^+$ ion effect, voltages of 5 kV and 2 kV were applied. The final thickness of the TEM sample was less than 100 nm. The in-situ STEM measurements were performed on a JEOL ARM300F TEM operated at 300 keV under 0, 2-50 V, and –(2-50) V bias using a Keithley 2400. During in-situ STEM measurement, a positive bias was constantly applied to the sample to ensure complete polarization switching. Bright-field in-situ TEM images of (111) planes of 3C-SiC under +4 V and −4 V bias were recorded on a JEOL ARM300F TEM operated at 300 keV using a Keithley 2400, with *g* slightly deviated from [110] by about 5° under a two-beam condition. The STEM images were recorded from more than five different regions to ensure reproducibility. After applying a negative bias to the sample, the bias was held for 5 min before collecting the STEM images to ensure complete polarization switching. The in-situ STEM measurements were then conducted following the same procedure as for the positive bias. Si atoms were clearly visible in the HAADF-STEM images and ADF-STEM images. The C(N) atoms were only clearly visible in the ADF-STEM images. The SAED patterns were recorded under 0 V, +4 V, and −4 V bias. For comparison, lightly N-doped 3C-SiC epitaxial films were also prepared by a standard FIB lift-off process (FEI SCIOS 2 using $Ga^+$ ions) onto a copper grid. The HAADF-STEM pattern was recorded on a JEOL ARM300F TEM operated at 300 keV.

Hall measurements were conducted on a physical property measurement system (PPMS, Quantum Design) at 2−300 K via the standard six-wire method. The wafer resistivity mapping was conducted using an NC-80MAP non-contact mapping system (NAPSON Corporation, Japan) based on the eddy current method. The carrier concentrations for the epitaxial 3C-SiC films were obtained from the *C*-*V* results obtained with a Hg-CV Analyzer (MCV-530L).

The first-principles calculations were carried out with the density functional theory (DFT) implemented in the Vienna ab initio simulation package (VASP)[58]. We adopted the generalized gradient approximation (GGA) in the form of the Perdew-Burke-Ernzerh (PBE)[59] for the exchange-correlation potentials. The projector-augmented-wave (PAW)[60] pseudopotentials were used with a plane wave energy of 500 eV. A $\Gamma$-centered Monkhorst-Pack[61] Brillouin zone sampling grid with a resolution of $0.02\times2\pi$ $Å^{-1}$ was applied. Atomic positions and lattice parameters were relaxed until all the forces on the ions were less than $10^{-2}$ eV·Å$^{-1}$. The self-consistent field procedure was considered converged when the energy difference between two consecutive cycles was lower than $10^{-6}$ eV. The IRVSP code[62] was used to get the irreducible representations. The Fermi Surfer code[63] was used to visualize the Fermi surface.

**Pseudo Jahn–Teller effect**

Substitutional N doping introduces free charge carriers and drives metallization in cubic 3C-SiC[23, 25]. The conduction band minimum (CBM) resides at three equivalent *X* high-symmetry points within the face-centered cubic Brillouin zone, forming triple-degenerate conduction valleys. Valley degeneracy enables abundant X↔X' intervalley electron hopping, which satisfies the momentum conservation relation $k_{x\prime} = k_x + q_{phonon} + \tau$ and acts as the dominant momentum-exchange channel for electron-phonon coupling (EPC)[64, 65]. The $k_x$ and $k_{x\prime}$ are the initial and final electron wavevectors, $q_{phonon}$ is the wavevector of the phonon, and $\tau$ is the reciprocal lattice vector accounting for Umklapp scattering.

The microscopic origin of the observed structural phase transition lies in the pseudo Jahn-Teller effect (PJTE)[26,66], whose active pseudo-degenerate electronic configuration is established through a symmetry-specific mechanism at the *Γ* point. In pristine 3C-SiC, crystal-field splitting breaks the triply degenerate valence band maximum ($T_2$) into a nondegenerate $A_1$ singlet and a doubly degenerate $E$ manifold. The highly electronegative substitutional N introduces a deep impurity level of $A_1$ symmetry, which hybridizes with and repels the existing valence $A_1$ subband. This interaction drives a substantial downward energy shift of the occupied $A_1$ state ($\Delta E_1 < 0$), alleviating intrinsic tetrahedral chemical strain and providing an initial electronic stabilization. Consequently, this occupied $A_1$ state and the empty, triply degenerate *X*-valley conduction states are brought into sufficient energetic proximity to form the active pseudo-degenerate electronic pair required for the PJTE. The vibronic coupling between these pseudo-degenerate states, mediated by a specific polar vibrational mode, drives a spontaneous symmetry-breaking lattice distortion from cubic $F\bar{4}3m$ to rhombohedral $R3m$, which is energetically favored.

Following this symmetry-lowering structural transition, the electronic structure undergoes pronounced many-body renormalizations. The large momentum transfers inherent to intervalley hopping lead to the softening and red-shift of transverse optical (TO) phonons to ~784 cm$^{-1}$. The softened phonon modes exert reciprocal back-coupling to conduction electrons via the electron-phonon self-energy. While the band bottom of each *X* valley remains nearly unchanged, the dispersions of unoccupied $\sigma^*$-antibonding $X$ manifolds are globally broadened and flattened[63,64], a direct consequence of triplet valley degeneracy and intense intervalley scattering.

This renormalization of the conduction band landscape drastically increases the density of states (DOS) of unoccupied states in the low-energy window. Within the framework of many-body perturbation theory, this elevated empty conduction DOS induces a downward orbital level repulsion on the occupied valence bands, leading to a collective reduction in the kinetic energy of valence electrons as follows: $\Delta E_2$ =($E_{F_1}$-$E_{F_2}$)<0.

Taken together, the PJTE-driven structural phase transition is cooperatively stabilized by two distinct energy-lowering channels: the initial electronic reorganization ($\Delta E_1$) and the subsequent many-body valence band downshift ($\Delta E_2$). The total energy reduction $\Delta E = \Delta E_1 + \Delta E_2$ ( $E_{F_1}$ - $E_{F_2}$ ) renders the $R3m$ phase thermodynamically favorable.

## Thomas-Fermi screening length calculations

The Thomas-Fermi screening length for a 3D degenerate electron gas is given by (*9*):

$$\lambda_{TF} = \sqrt{\frac{\varepsilon_0 \varepsilon_r}{e^2 D(E_f)}} \quad (1)$$

where $\varepsilon_0$ is the vacuum permittivity, $\varepsilon_r$ is the relative permittivity, $e$ is the elementary charge, and $D(E_f)$ is the density of states at the Fermi level. For a free electron gas, $D(E_f)$ is expressed as[67]:

$$D(E_f) = \frac{3\mathrm{n}}{2E_F} \quad (2)$$

with $n$ the electron concentration and $E_f$ the Fermi energy. Using the free-electron model, $E_f$ is calculated as[68]:

$$E_f = \frac{\hbar^2}{2m_e}(3\pi^2 n)^{2/3} \quad 3)$$

where $\hbar$ is the reduced Planck constant and $m_e$ is the free electron mass, $n = 2.9 \times 10^{20}\ cm^{-3}$, $\varepsilon_r = 9.7$. Finally, the calculated Thomas–Fermi screening length is: $\lambda_{TF} = 1.4\ nm$. This value is larger than the lattice constant of N-doped 3C-SiC ($c$ = 0. 7552(4) nm) and significantly larger than the Si-C bond length (~0.189 nm). Consequently, free carriers are unable to fully screen the local bond dipoles on the atomic scale, which provides a necessary condition for the stabilization of ferroelectric polarization in this metallic system.

**Coercive field**

The metallic nature of heavily N-doped 3C-SiC single crystals renders conventional polarization-electric field (*P*-*E*) hysteresis loop measurements infeasible [69,70]. To address this limitation, we estimated the coercive field to be lower than 2.5 kV/cm based on the PFM results. The coercive field is lower than those of conventional ferroelectrics such as $BaTiO_3$ (~5 kV/cm)[38] and congruent lithium niobate (~20 kV/cm)[71], PZT ceramics (~50 kV/cm)[48,72], $HfO_2$-based ferroelectrics (~0.1-5 MV/cm)[73-76], emerging two-dimensional ferroelectrics such as $Bi_2SeO_5$ (2.3-11 kV/cm)[47,77-78], and AlN-based ferroelectrics (2.9-17 MV/cm)[28,29,79-81].

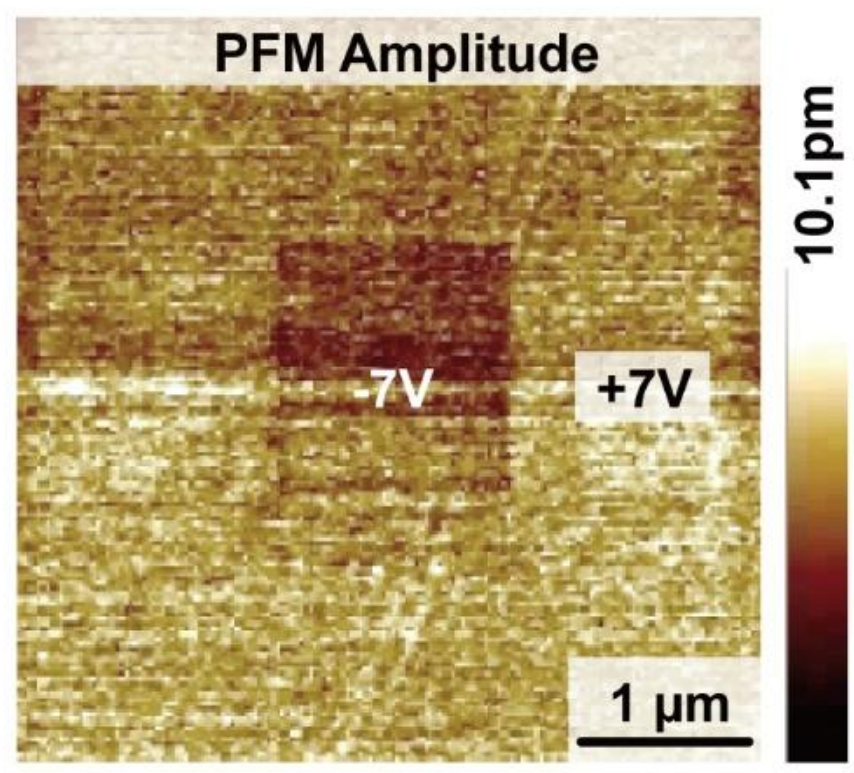

**Supplementary Fig. 1 | Ferroelectric behaviors of heavily N-doped 3C-SiC verified by PFM.** PFM amplitude image after box-in-box writing with +7 V for a 3×3 μm region and −7 V for a 1×1 μm region by applying a tip bias. Well-defined PFM amplitude contrast is obviously seen, showing the remanent polarization states and thus the switchable polarization.

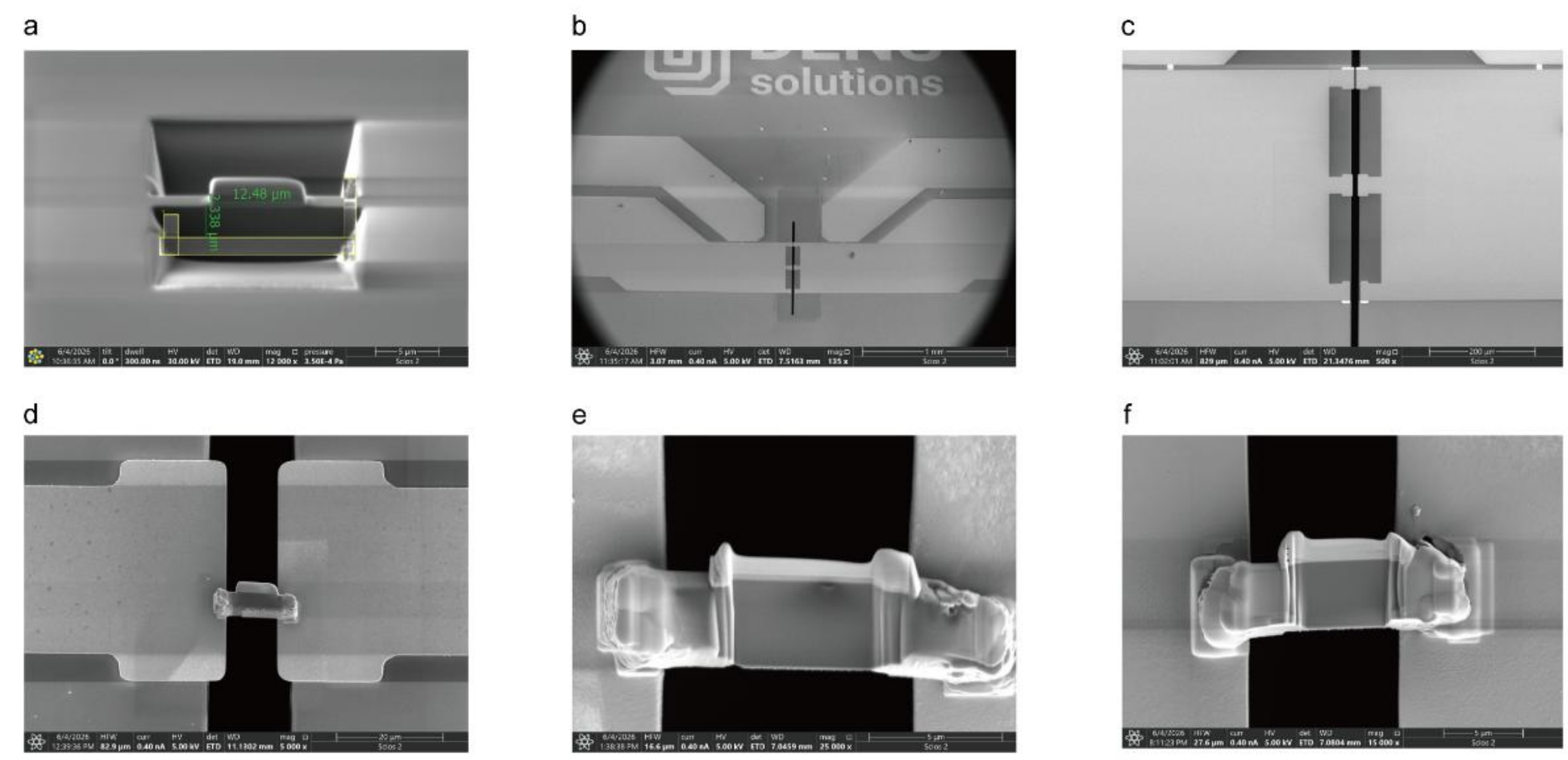

**Supplementary Fig. 2 | Stepwise SEM images documenting the focused ion beam (FIB) fabrication of a lamella for in-situ TEM and STEM measurements. a,** Cross-sectional SEM micrograph of the pre-thinned specimen after FIB milling, with the marked lateral length (12.48 μm) and thickness (2.338 μm) of the target lamella indicated. **b,** Low-magnification overview of the commercial electrical biasing chip platform. **c,** Medium-magnification SEM view of the chip's central electrode window gap. **d,** SEM image of the narrow gap between the two contact electrodes, where the lamella will be transferred and mounted. **e,** High-magnification SEM image of the lifted-out lamella after being welded onto the chip electrodes. **f,** Final TEM lamella ready for in-situ STEM measurements. A protective Pt layer covers the central sample region, and both ends of the specimen are electrically connected to the chip electrodes to enable external voltage biasing. All SEM images were acquired using a dual-beam FIB-SEM system.

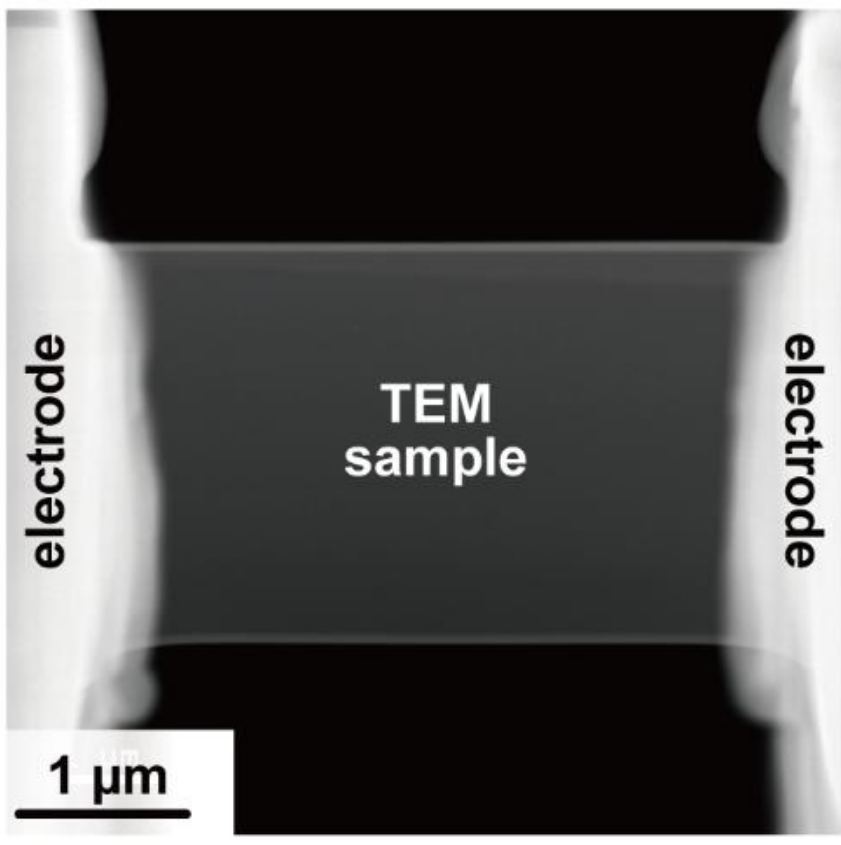

**Supplementary Fig. 3 | The TEM image of the sample for the in-situ atomic-scale observation of electric-field-induced polarization switching in heavily N-doped 3C-SiC single crystals.**

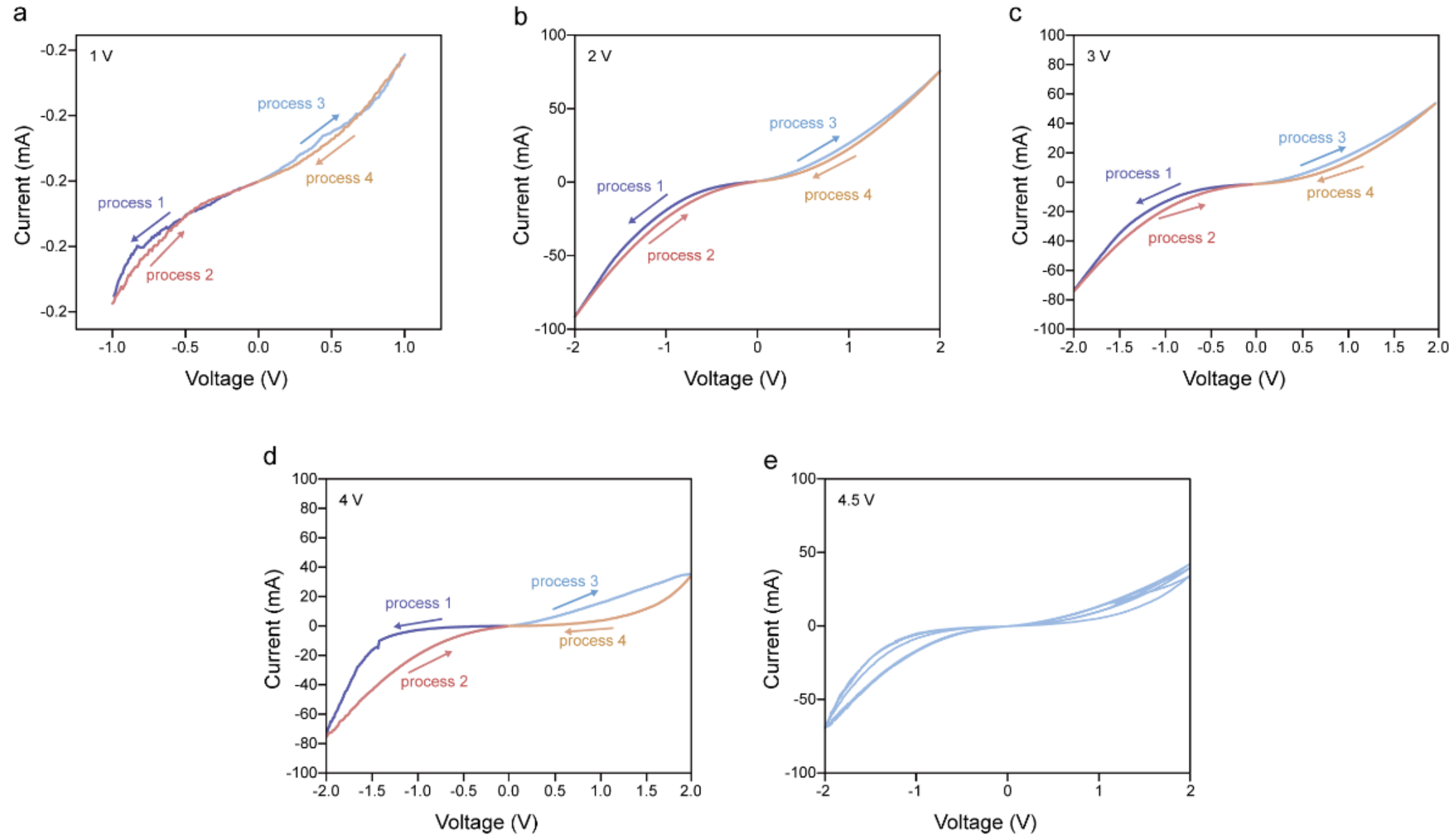

**Supplementary Fig. 4 | I-V curves of ferroelectric tunneling junctions fabricated on heavily N-doped 3C-SiC single crystals, measured after pre-polarization at different biases of (a) 1 V, (b) 2 V, (c) 3 V, (d) 4 V, (e) 4.5 V (three times).** The pre-polarizing voltage is removed prior to each measurement. Four distinct processes are seen: process 1 (negative sweep, black), process 2 (negative-to-zero return, red), process 3 (positive sweep, blue), and process 4 (positive-to-zero return, cyan). The hysteresis window systematically widens with increasing pre-polarization voltage, driven by enhanced field-induced ferroelectric polarization switching. Repeated I-V curves measured after 4.5 V pre-polarization, verifying stable and reproducible conductance switching of FTJs.

**Supplementary Table 1 |** Carrier concentration measured at nine random points on epitaxial 3C-SiC films.

| Points | Carrier concentration ($10^{15}$ $cm^{-3}$) |
|---|---|
| 1 | 8.29 |
| 2 | 8.60 |
| 3 | 8.46 |
| 4 | 8.64 |
| 5 | 8.36 |
| 6 | 8.06 |
| 7 | 8.53 |
| 8 | 8.69 |
| 9 | 9.65 |

**Supplementary Table 2 |** Crystallographic parameters for *R*3*m* obtained by single crystal XRD results.

| **Formula** | **$SiC_{0.994}N_{0.006}$** |
|---|---|
| Formula per unit cell | 1 |
| Space group | *R*3*m* |
| Space group number | 160 |
| $a$=$b$/Å | 3.0849(1) |
| $c$/Å | 7.552(4) |
| $\alpha$/° | 90 |
| $\beta$/° | 90 |
| $\gamma$/° | 120 |
| Unit-cell volume/Å$^3$ | 62.267(4) |

**Supplementary Table 3 |** Comparison of coercive voltage and Curie temperature with other ferroelectric materials.

| Materials | Thickness (nm) | Coercive voltage (V) | Curie temperature (°C) | Ref. |
|---|---|---|---|---|
| **N-doped 3C-SiC single crystals** | **350000** | **0.4** | **above 599** | **This work** |
| $BaTiO_3$ | single crystal | 1-50 V | ~120 | 30 |
| $BaTiO_3$ | 25 | ~0.1 | / | 38 |
| $PbTiO_3$ | 17 | 2.32 | ~490 | 87 |
| PZT | 280 | 0.3 | 230–490 | 88 |
| $LiOsO_3$ | | / | -133 | 3 |
| $(SnSe)_{1.16}(NbSe_2)$ | | / | ~110 | 6 |
| $WTe_2$ | | / | ~67 | 5 |
| $Al_{0.68}Sc_{0.32}N$ | 45 | / | Above 600 | 28 |
| $Bi_2O_2Se$ | 1.1 | ~0.8-0.9 V | 607 | 47 |
| $Hf(Zr)_{1+x}O_2$ | 12.1 | ~0.8 V | / | 89 |

**Supplementary Table 4 |** Comparisons for the performances of typical FTJs.

| Material | Response speed | Cycles | Retention duration | Ref. |
|---|---|---|---|---|
| $Ag_2S$ | NA | 60 | 500 s | 45 |
| BTO | 10 ns | 900 | NA | 90 |
| HZO | 100 ns | 1000 | NA | 91 |
| HZO | 10000 ns | 1E7 | 3.15E8 s | 92 |
| HZO | 36 ms | 1E5 | NA | 93 |
| BSO | 300 ps | 5E9 | 3.15E8 s | 94 |
| BTO | 600 ps | E8-E9 | 3.15E9 s | 95 |
| **N-doped 3C-SiC single crystals** | **50 ns** | | 3.15E9 s | **This work** |

**Supplementary Table 5 |** Benchmark of key properties for representative ferroelectric/polar metals.

| Materials | Lateral size (cm) | ρ (mΩ·cm) | On/off ratio | In-situ TEM | Switchable FE | Ref. |
|---|---|---|---|---|---|---|
| **N-doped 3C-SiC single crystals** | **15** | **0.52** | **20** | **Yes** | **Yes** | **This work** |
| $LiOsO_3$ | 0.69 | 1.6 | N/A | No | No | 3 |
| $WTe_2$ | 0.6-0.8 (2D flake) | 0.4 | 10-100 | No | Yes | 5 |
| $(SnSe)_{1.16}(NbSe_2)$ | 0.1 | 30 | N/A | No | Yes | 6 |
| $SrRuO_3$ | 1 cm (thin film) | 0.1 | 10-50 | No | Yes | 7 |
| $Cd_2Re_2O_7$ | / | / | ~0.1 (est.) | No | No | 12 |
| $PtBi_2$ | / | / | / | No | / | 82 |
| Thin film $ANiO_3$ | / | 1-10 (2.2 nm) | / | No | / | 83 |
| $TaNiTe_5$ | ~$10^{-4}$ (few-μm flake) | / | / | No | Yes | 84 |
| $Ca_3Co_3O_8$ | 0.05 (single crystal) | ~5 | / | No | No (polar magnet) | 85 |
| $Mg_3Cl_7$ | <0.01 (high-pressure) | / | / | No | No (polar metal) | 86 |